\documentclass[twocolumn,tighten,twocolappendix]{aastex7}

\usepackage{graphicx} 
\usepackage{graphics}
\usepackage{epsfig}
\usepackage{natbib}
\usepackage{mathtools}
\usepackage{caption}

\begin{document}

\newcommand{\NRAO}{\affiliation{National Radio Astronomy Observatory, 520 Edgemont Road, Charlottesville, VA 22903, USA}}
\newcommand{\UVa}{\affiliation{Department of Astronomy, University of Virginia, Charlottesville, VA 22904}}

\title{The Star Formation in Radio Survey: Adding 90\,GHz Data to 3\textendash33\,GHz Observations of Star-forming Regions in Nearby Galaxies}
\shorttitle{ SFRS: 3\textendash90\,GHz Radio 
}
\shortauthors{Dignan et al.}

\author[0009-0000-8877-135X]{Anna Dignan}
\NRAO \UVa \email{adignan@nrao.edu}
\author[0000-0001-7089-7325]{Eric J. Murphy}
\NRAO \email{emurphy@nrao.edu}
\author[0000-0002-8472-836X]{Brian Mason}
\NRAO \email{bmason@nrao.edu}
\author[0000-0002-1185-2810]{Cosima Eibensteiner} \email{ceibenst@nrao.edu}
\altaffiliation{Jansky Fellow of the National Radio Astronomy Observatory}
\NRAO
\author[0000-0001-7449-4638]{Brandon S. Hensley} \email{brandon.s.hensley@jpl.nasa.gov}
\affiliation{Jet Propulsion Laboratory, California Institute of Technology, 4800 Oak Grove Drive, Pasadena, CA 91109, USA}
\author[0000-0002-2640-5917]{Eric F. Jim\'enez-Andrade} \email{xm.manu.ayri@zenemij.e}
\affiliation{Instituto de Radioastronom\'ia y Astrof\'isica, Universidad Nacional Aut\'onoma de M\'exico, Antigua Carretera a P\'atzcuaro \# 8701, Ex-Hda. San Jos\'e de la Huerta, Morelia, Michoac\'an, M\'exico C.P. 58089}
\author[0000-0002-1000-6081]{Sean T. Linden} \email{seanlinden@arizona.edu}
\affiliation{Steward Observatory, University of Arizona, 933 N Cherry Avenue, Tucson, AZ 85721, USA}
\author[0000-0002-1940-4289]{Simon R. Dicker} \email{sdicker@nrao.edu}
\affiliation{Department of Physics and Astronomy, University of Pennsylvania, 209 S. 33rd St., Philadelphia, PA 19014, USA}
\author[0000-0001-9584-2531]{Dillon Z. Dong} \email{ddong@nrao.edu}
\affiliation{National Radio Astronomy Observatory, P.O. Box 0, Socorro, NM 87801, USA}
\author[0000-0003-3168-5922]{Emmanuel Momjian} \email{emomjian@nrao.edu}
\affiliation{National Radio Astronomy Observatory, 1011 Lopezville Rd.,
Socorro, NM 87801, USA}
\author[0000-0001-5725-0359]{Charles E.\ Romero} \email{charles.romero@gmail.com}
\affiliation{Department of Physics and Astronomy, University of Pennsylvania, 209 South 33rd Street, Philadelphia, PA, 19104, USA}
\author[0000-0002-3933-7677]{Eva Schinnerer} \email{schinner@mpia.de}
\affiliation{Max Planck Institute for Astronomy, Königstuhl 17, 69117, Heidelberg, Germany}
\author[0000-0003-4625-2951]{Jean L. Turner} \email{turner@astro.ucla.edu}
\affiliation{Department of Physics and Astronomy, UCLA, Los Angeles, CA 90095, USA}

\date{\today}

\begin{abstract} 
\noindent 
We present 90\,GHz continuum imaging of 119 star-forming regions in 30 nearby galaxies observed with MUSTANG-2 on the Robert C. Byrd Green Bank Telescope as part of the Star Formation in Radio Survey. 
The 90\,GHz data were combined with 3, 15, and 33\,GHz data taken previously by the Karl G. Jansky Very Large Array to decompose radio spectra on $\approx$0.8\,kpc scales into their synchrotron, free-free, and thermal dust emission components. 
This was done using three scenarios: 
(i) a power law fit from 3 to 33\,GHz, 
(ii) Markov Chain Monte Carlo (MCMC) fitting from 3 to 90\,GHz with a thermal dust component, and 
(iii) MCMC fitting from 3 to 33\,GHz without a thermal dust component. 
For these cases, we find a median thermal (free-free) emission fraction at 33\,GHz of (i) $88 \pm 2$\% with a scatter of 17\%, (ii) $76\pm 3$\% with a scatter of 25\%, and (iii) $84\pm 2$\% with a scatter of 18\%. 
From this we conclude that, on average, free-free, not thermal dust, remains the dominant emission component at 33\,GHz. 
 While scenario (ii) yields a thermal fraction that is $\approx$10\% larger than scenario (iii), this difference decreases to $\approx$5\% after AGN are removed. 
 Consequently, star formation rates measured with thermal fractions at 33\,GHz are only mildly biased high without 90\,GHz data for the spectral decomposition.  
Furthermore, a power law fit of data from 3 to 33\,GHz still provides a reliable estimate of the free-free emission at 33\,GHz.

\end{abstract}

\section{INTRODUCTION}
\label{INTRODUCTION}
Massive star formation shapes galaxy properties in our Universe through their production and dispersal of heavy elements into the interstellar medium (ISM), impact on galaxy structure and evolution, and influence on planetary systems \citep{McKee2007,Kennicutt2012,Klessen2016}. Radio continuum emission from massive star-forming galaxies is powered by a combination of non-thermal synchrotron radiation from ultrarelativistic electrons and thermal free-free radiation generated in H\,II regions. The strength of the free-free component is directly proportional to the total amount of ionizing radiation generated by O and B type stars within these H\,II regions. These stars end their lives as Type II core-collapse supernovae, in turn creating remnants that accelerate cosmic ray electrons into the magnetized ISM of galaxies, producing synchrotron emission \citep[]{Condon1992,Koyama1995}. 

Previous studies have shown that at lower frequencies ($\sim$1\,GHz), the steep ($\alpha^{\mathrm{NT}}\sim-0.8$) synchrotron emission dominates the radio spectrum, while the free-free emission ($\alpha^{\mathrm{T}}\sim0.1$) becomes dominant at higher frequencies (${\sim}$30\,GHz) \citep[e.g.][]{Condon1992,Murphy2011,Tabatabaei2017}. These studies then derived star formation rate calibrations based on the decomposition of radio emission into their free-free and synchrotron emission components. Therefore, radio spectra, characterized by a simple power law, can be used as an extinction-free way to calculate the current number of massive ($\geq$ 8\(M_\odot\)) short-lived ($\leq$ 10 Myr) stars in a galaxy. 

Prior investigations have either focused on large samples at low (${\lesssim}$ 30\,GHz) frequencies \citep[e.g.][]{Karim2011,Rosero2016,Muxlow2020} or small sample sizes at high (90\,GHz) frequencies \citep[e.g.][]{Tabatabaei2007,Murphy2015,Nikolic2012,Chen2023}. This paper presents data taken as part of the Star Formation in Radio Survey \citep[SFRS;][]{Murphy2012,Murphy2018,Linden2020}, the goal of which is to study a large sample of radio spectra from nearby galaxies to better calibrate indicators of massive star formation. The SFRS aims to to bridge the gap between the two previous approaches by studying an extensive selection of galaxies, covering a range of properties and interstellar medium conditions, at multiple frequencies spanning a wide range in the radio regime. The survey's multifrequency observations add robust constraints to values for the spectral indices, improving empirical star formation rate calibrations in the radio overall. 

Initial SFRS observations were taken using the Westerbork Synthesis Radio Telescope and Green Bank Telescope (GBT) at 1.7\textendash33\,GHz for 53 nuclear and extranuclear star-forming regions on $\approx$1 kiloparsec scales \citep{Murphy2012}. They found that the measured thermal fraction at 33\,GHz for star-forming regions was dependent on physical resolution, but that additional data at different frequencies would be necessary to interpret the processes behind this trend and increase the accuracy of star formation rate estimates. \citet{Murphy2018} extended this study into a multifrequency Karl G. Jansky Very Large Array (VLA) campaign, which imaged 112 nuclear and extranuclear star-forming regions at 33\,GHz at an angular resolution of 2\arcsec and combined these observations with corresponding H$\alpha$ and 24\,$\mu$m photometry. In addition to analyzing the H$\alpha$ to 33\,GHz and 24\,$\mu$m to 33\,GHz flux density ratios, they also confirmed the presence of anomalous microwave emission (AME) in one of the imaged star-forming regions, making it the second known detection of extragalactic AME \citep{ame}. Most recently, \citet{Linden2020} obtained 3 and 15\,GHz imaging as part of the same campaign and included the previous 33\,GHz observations in measuring the spectral indices and thermal emission fractions at 33\,GHz. They found an increased scatter in measured spectral indices and thermal fractions as the galactocentric radius decreased, i.e., approached the galaxy's nucleus. They also identified 33 regions whose increasing emission from 15 to 33\,GHz indicated the potential influence of AME, noting that additional observations at higher frequency ($\geq$ 40\,GHz) would be required to confirm the presence of any AME. 

The new 90\,GHz data presented in this paper builds on these previous works, as they are sensitive to the part of the radio spectrum that is expected to be predominantly powered by free-free and thermal dust emission. When combined with existing 3, 15, and 33\,GHz data, it becomes possible to properly separate this free-free emission from other emission mechanisms that produce the observed radio spectrum. The resulting library of radio spectra at sub-kpc scales will be useful for understanding the nature of the radio emission from objects at high redshift, where there is currently less data available for comparison. 

This paper is organized as follows. In Section \ref{section:SAMPLE AND DATA}, we describe our sample selection, data reduction, and imaging procedure for the 90\,GHz GBT data. In Section \ref{section:PHOTOMETRY AND ANALYSIS}, we discuss the analysis procedures used. We present our results in Section \ref{section:RESULTS} and discuss them further in Section \ref{section:DISCUSSION}.  We summarize our conclusions in Section \ref{section:CONCLUSIONS}. 

Throughout this paper we report the scatter of relationships using scaled median absolute deviations (MADs) instead of standard deviations, as this statistic is less influenced by any outliers in data. The scaled MAD (MADN) is defined as the MAD divided by 0.6745. Unless otherwise noted, thermal fractions (i.e., fraction of free-free to total radio emission) are evaluated at 33\,GHz. 

\section{SAMPLE AND DATA}
\label{section:SAMPLE AND DATA}

In this section, we describe the sample selection and present the VLA and new GBT observations along with our data reduction and imaging procedures.

\subsection{Sample Selection}
The SFRS is made up of 56 nearby ($d_{L}$ $<$ 30 Mpc) galaxies sourced from the Spitzer Infrared Nearby Galaxies Survey \citep[SINGS;][]{Kennicutt2003} and Key Insights on Nearby Galaxies: a Far-Infrared Survey with Herschel \citep[KINGFISH;][]{Kennicutt2011} legacy programs (Table \ref{tab:properties}). The SINGS and KINGFISH galaxies represent the ISM conditions in the local universe and span the full range of morphological types, luminosities, and far infrared/optical luminosity ratios. The nuclear and extranuclear star-forming regions of the SFRS sample were chosen based on pre-existing mid-infrared spectral mappings collected by the InfraRed Spectrograph (IRS) instrument on board the Spitzer Space Telescope \citep{irs} and far-infrared spectral mappings taken by the Photodetector Array Camera and Spectrometer (PACS) instrument on board the Herschel Space Observatory \citep{pacs}. Combined, these spectral mappings cover the principal atomic ISM cooling lines: [O\,I] 63\,$\mu$m, [O\,III] 88\,$\mu$m, [N\,II] 122 and 205\,$\mu$m, and [C\,II] 158\,$\mu$m. NGC\,5194 was part of the SINGS sample but not formally included in the KINGFISH sample. Additionally, NGC\,2146 was included in the KINGFISH sample and has Spitzer data but was not part of SINGS.

Radio data used in this analysis include a total of 
119 regions located in 30 of the SFRS galaxies that are observable with both the VLA and GBT. 
Nuclear regions are defined as having galactocentric radii $r_{\rm G} < $ 250 pc, with extranuclear regions having $r_{\rm G} \geq $ 250 pc. 
Of the 119 star-forming regions, 23 are nuclear regions and 96 are extranuclear regions. 
Galaxy morphological types, luminosity distances, optically defined nuclear types, diameters ($\mathrm{D_{25}}$), inclinations (\textit{i}), and position angles (P.A.) for each source are given in Table \ref{tab:properties} and described in detail in \citet{Murphy2018}.
The coordinates given in Table \ref{tab:gbtdata} correspond to the pointing centers of the VLA observations (see Section \ref{subsection:VLA}). 

\begin{deluxetable*}{c|cccccc}
\tablecaption{Galaxy Properties\label{tab:properties}}
\tablehead{
\colhead{Galaxy} & 
\colhead{Type$^{a}$} &
\colhead{Dist.$^{b}$} & 
\colhead{Nuc. Type$^{c}$} & 
\colhead{{D$_{25}$}$^{a}$} & 
\colhead{\textit{i}} & 
\colhead{P.A.$^{a}$} \\
\colhead{} &
\colhead{} &
\colhead{(Mpc)} &
\colhead{} &
\colhead{(arcmin)} &
\colhead{($^{\circ}$)} &
\colhead{($^{\circ}$)}
}
\startdata
NGC\,0337 & SBd & 19.3 & SF & $2.9 \times 1.8$ & 52 & 130 \\
NGC\,0628 & SAc & 7.2 & ... & $10.5 \times 9.5$ & 25 & 25 \\
NGC\,0925 & SABd & 9.12 & SF & $10.5 \times 5.9$ & 57 & 102 \\
NGC\,2146 & Sbab & 17.2 & SF(*) & $6.0 \times 3.4$ & 56 & 57 \\
NGC\,2798 & SBa & 25.8 & SF/AGN & $2.6 \times 1.0$ & 70 & 160 \\
NGC\,2841 & SAb & 14.1 & AGN & $8.1 \times 3.5$ & 66 & 147 \\
NGC\,3049 & SBab & 19.2 & SF & $2.2 \times 1.4$ & 51 & 25 \\
NGC\,3190 & SAap & 19.3 & AGN(*) & $4.4 \times 1.5$ & 73 & 125 \\
NGC\,3184 & SABcd & 11.7 & SF & $7.4 \times 6.9$ & 21 & 135 \\
NGC\,3198 & SBc & 14.1 & SF & $8.5 \times 3.3$ & 68 & 35 \\
NGC\,3351 & SBb & 9.33 & SF & $7.4 \times 5.0$ & 48 & 13 \\
NGC\,3521 & SABbc & 11.2 & SF/AGN(*) & $11.0 \times 5.1$ & 63 & 163 \\
NGC\,3627 & SABb & 9.38 & AGN & $9.1 \times 4.2$ & 64 & 173 \\
NGC\,3773 & SA0 & 12.4 & SF & $1.2 \times 1.0$ & 33 & 165 \\
NGC\,3938 & SAc & 17.9 & SF(*) & $5.4 \times 4.9$ & 25 & 29$^{d}$ \\
NGC\,4254 & SAc & 14.4 & SF/AGN & $5.4 \times 4.7$ & 30 & 24$^{d}$ \\
NGC\,4321 & SABbc & 14.3 & AGN & $7.4 \times 6.3$ & 32 & 30 \\
NGC\,4536 & SABbc & 14.5 & SF/AGN & $7.6 \times 3.2$ & 66 & 130 \\
NGC\,4559 & SABcd & 6.98 & SF & $10.7 \times 4.4$ & 67 & 150 \\
NGC\,4569 & SABab & 9.86 & AGN & $9.5 \times 4.4$ & 64 & 23 \\
NGC\,4579 & SABb & 16.4 & AGN & $5.9 \times 4.7$ & 37 & 95 \\
NGC\,4594 & SAa & 9.08 & AGN & $8.7 \times 3.5$ & 69 & 90 \\
NGC\,4625 & SABmp & 9.3 & SF & $2.2 \times 1.9$ & 31 & 28$^{d}$ \\
NGC\,4631 & SBd & 7.62 & SF(*) & $15.5 \times 2.7$ & 83 & 86 \\
NGC\,4725 & SABab & 11.9 & AGN & $10.7 \times 7.6$ & 45 & 35 \\
NGC\,5194 & SABbcp & 7.62 & AGN & $11.2 \times 6.9$ & 53 & 163 \\
NGC\,5474 & SAcd & 6.8 & SF(*) & $4.8 \times 4.3$ & 27 & 98$^{d}$ \\
NGC\,5713 & SABbcp & 21.4 & SF & $2.8 \times 2.5$ & 27 & 10 \\
NGC\,6946 & SABcd & 6.8 & SF & $11.5 \times 9.8$ & 32 & 53$^{d}$ \\
NGC\,7331 & SAb & 14.5 & AGN & $10.5 \times 3.7$ & 72 & 171 \\
\enddata
\tablenotetext{a}{Morphological types, apparent major isophotal diameters, and position angles are taken from the Third Reference Catalog of Bright Galaxies \citep[RC3;][]{RC3}.}
\tablenotetext{b}{Redshift-independent distances are taken from Table 1 of \citet{Kennicutt2011}, except for the non-KINGFISH galaxy NGC\,5194 \citep{Ciardullo2002}.}
\tablenotetext{c}{Nuclear type is identified by optical spectroscopy. SF=star-forming; AGN=synchrotron emission. Nuclear types without an asterisk (*) are taken from Table 5 of \citet{Moustakas2010}, while those with one are taken from Table 4 of \citet{Lonsdale1987}.}
\tablenotetext{d}{Position angles are taken from \citet{Jarrett2003}.}
\end{deluxetable*}

\subsection{GBT Observations and Data Reduction}

Observations of the 30 targets were taken at 90\,GHz using MUSTANG-2 \citep{MUSTANG2}, a bolometer camera operating on the GBT.
The resulting angular resolution of these observations is on average for all sources $\sim$\,10\arcsec. 
Targets were chosen so they would be visible during portions of the year when weather conditions are suitable for high frequency observing with the GBT. Data were collected during two observing campaigns: April, October, November, and December 2019 (GBT/19A-284, PI: E. Murphy and GBT/19B-012, PI: E. Murphy) and February and March 2021 (GBT/21A-250, PI: E. Murphy). Due to poor weather conditions during initial observations, NGC\,5194 was re-observed using Director's Discretionary Time in April 2024 (GBT/24A-445; PI: Dignan). The typical integration time per target was 0.68\,hr, and we achieved a typical map noise of 81\,$\mu$Jy\,bm$^{-1}$. 
A customized, elliptical daisy scan pattern was developed in order to adequately cover each target. Specifically, the major and minor diameters and position angles given in Table \ref{tab:properties} were used to generate scan trajectories that fully and efficiently covered each galaxy. Data were calibrated and imaged using the standard MUSTANG-2 pipeline described in \citet{Romero2020,Romero2017}, with the absolute flux density scale determined by observations of Atacama Large Millimeter/submillimeter Array (ALMA) grid calibration sources \citep{Fomalont2014}. While not included in this analysis, the typical flux calibration uncertainty for MUSTANG-2 is $\sim10\%$ \citep{Romero2020}. The full width half maximum (FWHM) of the synthesized circular beam and the corresponding point-source and brightness temperature sensitivities for each image are given in Table \ref{tab:gbtdata}. 

\begin{deluxetable*}{c|cc|ccc}
\tablecaption{Target Positions and Imaging Characteristics at 90\,GHz\label{tab:gbtdata}} 
\tablewidth{0pt}
\tablehead{
\colhead{Galaxy} & 
\colhead{R.A. (J2000)} & 
\colhead{Decl. (J2000)} & 
\colhead{Synthesized Beam} & 
\colhead{$\sigma$} & 
\colhead{$\sigma_{{T}_{b}}$} \\
\colhead{} &
\colhead{(hh mm ss.ss)} &
\colhead{(dd mm ss.ss)} &
\colhead{} & 
\colhead{($\mu$Jy\,bm$^{-1}$)} & 
\colhead{($\mu${K})}
}
\startdata
NGC\,0337 & 00 59 50.30 & $-07$ 34 44.00 & 9.53$\arcsec$ & 84.31 & 139.42 \\
NGC\,0628 & 01 36 41.70 & 15 46 59.00 & 11.34$\arcsec$ & 59.58 & 69.59 \\
NGC\,0925 & 02 27 17.00 & 33 34 43.00 & 10.72$\arcsec$ & 154.21 & 201.29 \\
NGC\,2146 & 06 18 37.71 & 78 21 25.30 & 9.52$\arcsec$ & 93.70 & 155.31 \\
NGC\,2798 & 09 17 22.79 & 41 59 59.02 & 9.85$\arcsec$ & 72.02 & 111.53 \\
NGC\,2841 & 09 22 2.63 & 50 58 35.47 & 9.16$\arcsec$ & 89.79 & 160.82 \\
NGC\,3049 & 09 54 49.56 & 09 16 15.94 & 11.01$\arcsec$ & 80.81 & 100.09 \\
NGC\,3190 & 10 18 16.86 & 41 25 26.59 & 9.21$\arcsec$ & 91.37 & 161.59 \\
NGC\,3184 & 10 18 5.63 & 21 49 56.26 & 9.97$\arcsec$ & 79.20 & 119.61 \\
NGC\,3198 & 10 19 54.95 & 45 32 59.64 & 9.00$\arcsec$ & 94.46 & 174.93 \\
NGC\,3351 & 10 43 57.70 & 11 42 13.70 & 10.57$\arcsec$ & 65.38 & 87.82 \\
NGC\,3521 & 11 05 48.58 & $-$00 02 9.11 & 9.18$\arcsec$ & 90.56 & 161.41 \\
NGC\,3627 & 11 20 14.96 & 12 59 29.54 & 10.38$\arcsec$ & 488.16 & 679.75 \\
NGC\,3773 & 11 38 12.87 & 12 06 43.47 & 11.11$\arcsec$ & 75.08 & 91.30 \\
NGC\,3938 & 11 52 49.45 & 44 07 14.60 & 9.95$\arcsec$ & 91.94 & 139.54 \\
NGC\,4254 & 12 18 49.40 & 14 24 59.00 & 10.68$\arcsec$ & 56.16 & 73.89 \\
NGC\,4321 & 12 22 54.90 & 15 49 21.00 & 11.61$\arcsec$ & 99.78 & 111.18 \\
NGC\,4536 & 12 36 49.80 & 13 09 46.00 & 10.82$\arcsec$ & 98.35 & 126.07 \\
NGC\,4559 & 12 37 43.60 & 11 49 2.00 & 11.28$\arcsec$ & 89.94 & 106.15 \\
NGC\,4569 & 12 39 59.40 & $-$11 37 23.00 & 10.31$\arcsec$ & 67.70 & 95.54 \\
NGC\,4579 & 12 50 26.60 & 25 30 6.00 & 9.96$\arcsec$ & 62.96 & 95.27 \\
NGC\,4594 & 20 34 52.32 & 60 09 14.00 & 9.63$\arcsec$ & 90.73 & 146.92 \\
NGC\,4625 & 22 37 4.10 & 34 24 56.00 & 9.51$\arcsec$ & 88.08 & 146.21 \\
NGC\,4631 & 12 34 27.10 & 02 11 17.00 & 9.39$\arcsec$ & 87.19 & 148.32 \\
NGC\,4725 & 12 35 57.70 & 27 57 36.00 & 10.53$\arcsec$ & 86.95 & 117.64 \\
NGC\,5194 & 12 41 52.40 & 41 16 24.00 & 9.58$\arcsec$ & 342.43 & 559.94 \\
NGC\,5474 & 13 29 52.70 & 47 11 43.00 & 9.47$\arcsec$ & 151.42 & 253.47 \\
NGC\,5713 & 14 40 11.30 & $-$00 17 27.00 & 10.32$\arcsec$ & 97.00 & 136.79 \\
NGC\,6946 & 14 05 1.30 & 53 39 44.00 & 9.64$\arcsec$ & 141.01 & 227.99 \\
NGC\,7331 & 12 42 8.10 & 32 32 29.40 & 10.33$\arcsec$ & 64.70 & 91.06 \\
\enddata
\tablecomments{See \citet{Linden2020} for the 3 and 15\,GHz imaging characteristics and \citet{Murphy2018} for the 33\,GHz imaging characteristics.}
\end{deluxetable*}

\subsection{VLA Observations and Data Reduction}
\label{subsection:VLA}

Observations at 3, 15, and 33\,GHz were obtained using the VLA
at an angular resolution of $\sim$\,2\arcsec. 
Details on these observations, as well as data reduction and imaging characteristics, can be found in \citet{Murphy2018} for the 33\,GHz data and \citet{Linden2020} for the 3 and 15\,GHz data. 
To summarize, observations in the Ka-band (29\textendash37\,GHz) were taken during two VLA D-configuration cycles in November 2011 (11B-032) and March 2013 (13A-129); observations in the S-band (2\textendash4\,GHz) were taken during the 2013 VLA B-configuration cycle (13B-215), and observations in the Ku-band (12\textendash18\,GHz) were taken in November 2014 during the C configuration (13B-215). Data reduction followed standard calibration procedures using the VLA calibration pipeline built on Common Astronomy Software Applications \citep[CASA;][]{CASA} versions 4.6.0 and 4.7.0. After each initial pipeline run, the calibration tables and visibilities were manually inspected; any signs of instrumental issues (such as bad deformatters) and radio frequency interference were flagged. After flagging, the pipeline was rerun and this process was repeated until any further indicators of bad data could not be identified. Self-calibration was performed for NGC\,4594 and NGC\,4579, which showed phase and amplitude issues due to their bright AGN. The calibrated VLA measurement sets for each source were then imaged using the task \texttt{tclean} in CASA version 4.7.0. The deconvolver mode was set to multiscale multifrequency synthesis. The scales used were 0, 8, and 24 pixels, corresponding to different fractions of the FWHM of the point spread function. The typical flux calibration uncertainty of VLA data is $\sim3\%$ \citep{Perley2013} and, as with the GBT observations, is not included in this analysis.

After imaging the VLA data, a primary beam correction was applied using the CASA task \texttt{impbcor}. 
Each image was convolved to match the exact angular resolution of the corresponding GBT image ($\approx$10\arcsec on average for all sources) using the CASA task \texttt{imsmooth}. Lastly, the images were regridded to the same pixel scale (1.8\arcsec) as the corresponding GBT images using the CASA task \texttt{imregrid}. The resulting average rms values for the resolution-matched and regridded images at 3, 15, and 33\,GHz are 100, 96, and 68\,$\mu$Jy\,bm$^{-1}$, respectively.

\section{PHOTOMETRY AND ANALYSIS}
\label{section:PHOTOMETRY AND ANALYSIS}

In this section, we describe our procedure for aperture photometry for all four frequencies (i.e., 3, 15, 33, and 90\,GHz) at matched resolution. We also describe our method for calculating flux densities and decomposing our radio spectra into their individual emission components.

Aperture photometry was performed using the function \texttt{aperture\_photometry} from the Python package \texttt{photutils} \citep{photutils}. The source locations chosen for aperture photometry are the same as those given in Table~5 of \citet{Linden2020}. Sources are classified into 106 star-forming regions, one source associated with a potential supernova remnant, and 12 sources that may contain anomalous microwave emission based on the designations given in \citet{Linden2020}. In general, a region is classified as a likely AME candidate if its 33\,GHz emission is elevated relative to what is expected from lower frequency data using a two-component power law. An aperture diameter equal to twice the full width at half maximum of the corresponding GBT beam was used, resulting in a median aperture diameter of $\sim 0.8 \pm 0.01$\,kpc with a scaled median absolute deviation of $\approx 0.5$\,kpc. Additionally, annuli were used to calculate the median of the local background, with an inner diameter equal to 3 times the full width at half maximum of the corresponding GBT beam and an outer diameter equal to the sum of the inner and aperture diameters (i.e., 5 $\times$ FWHM). 
The local backgrounds contribute 1.2, 4.1, 0.2, and 9.9\% to the measured flux densities at 3, 15, 33, and 90\,GHz, respectively; these percentages were found by taking the median of the median flux density measured in the annulus divided by the integrated flux density measured in the aperture.  

\begin{figure*}
    \centering
\includegraphics[width=1.0\textwidth]{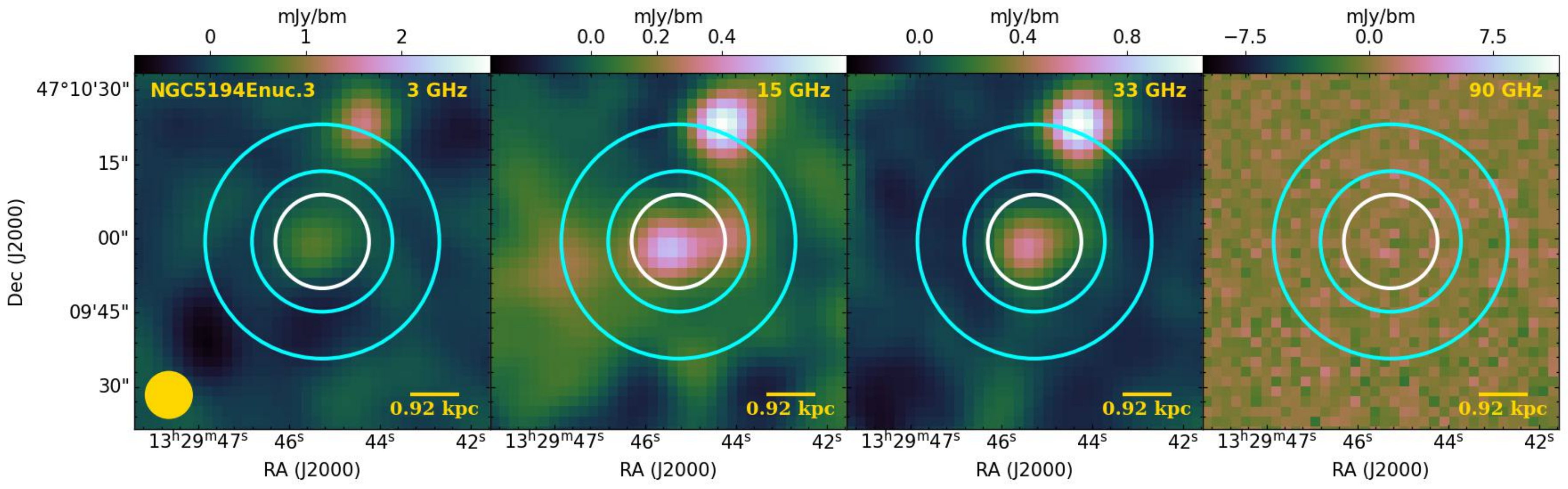}
    \caption{Images at each frequency for the extranuclear region NGC\,5194\,Enuc.\,3. The white circles correspond to the apertures used to perform photometry, while the cyan circles represent the annuli used to determine the local background. A corresponding version of this figure for all other sources is available in the online journal as a figure set (120 images).}
    \label{fig:photometry}
\end{figure*}

Figure \ref{fig:photometry} shows an example source, the extranuclear region NGC\,5194\,Enuc.\,3, at each frequency with the appropriate apertures and annuli. Corresponding versions of this figure for all of the sources in our sample are available in the online journal as a figure set (see Figure \ref{fig:photometry}). Aperture photometry results, as well as region locations, diameter of the aperture used for photometry ($d_{\mathrm{ap}}$), and galactocentric radius ($r_{\mathrm{G}}$) are given in Table \ref{tab:photometry}. We report all flux densities as measured, including non-detections (S/N $<$ 3).

Upon reviewing the data taken with the VLA for NGC\,2146, we concluded that 15\,GHz may have suffered from calibration issues. Although flux densities at 15\,GHz are reported in \citet{Linden2020}, we exclude them here. Additionally, we do not report flux densities for NGC\,4631\,Enuc.\,1 due to its location falling outside of the 90\,GHz map.

\noprint{\figsetstart}
\noprint{\figsetnum{1}}
\noprint{\figsettitle{Cutouts}}

\figsetgrpstart
\figsetgrpnum{1.1}
\figsetgrptitle{NGC5194Enuc.3}
\figsetplot{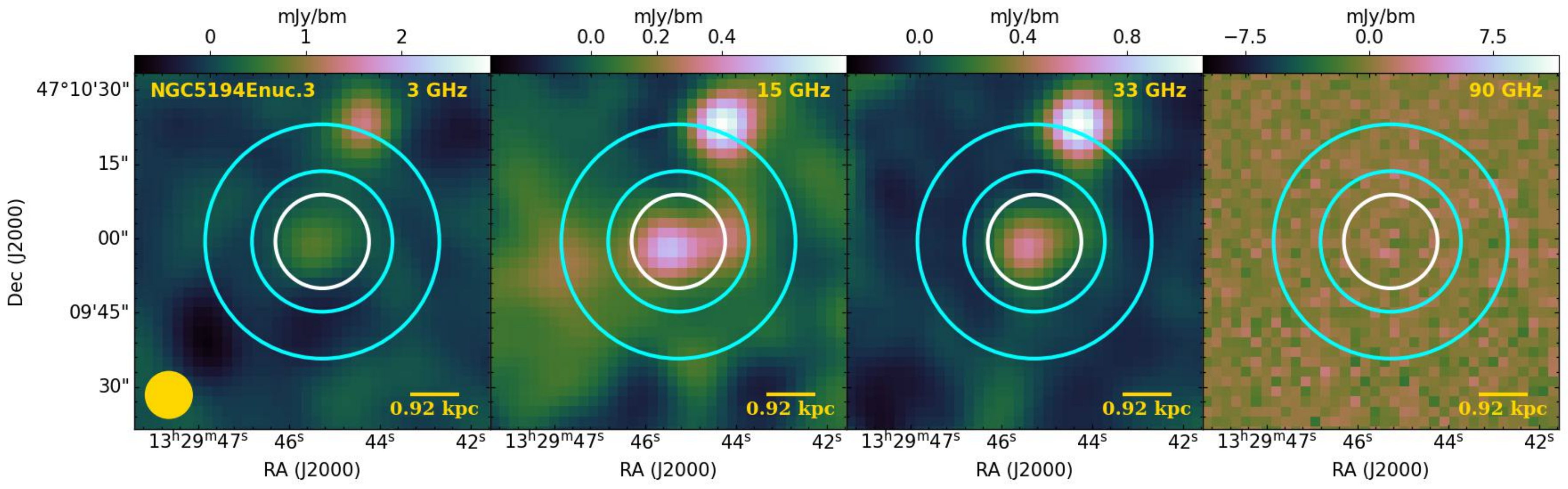}
\figsetgrpnote{Images at each frequency for the corresponding region.}
\figsetgrpend

\figsetend

\startlongtable
\begin{deluxetable*}{lcc|cccc|cc} 
\label{tab:photometry}
\tablecolumns{9}
\tablecaption{Region Photometry at 10\arcsec Angular Resolution}
\tablehead{
\colhead{Source ID} &
\colhead{R.A. (J2000)} &
\colhead{Decl. (J2000)} &
\colhead{$S_\mathrm{3\,GHz}$} &
\colhead{$S_\mathrm{15\,GHz}$} &
\colhead{$S_\mathrm{33\,GHz}$} &
\colhead{$S_\mathrm{90\,GHz}$} &
\colhead{$d_{\mathrm{ap}}$} &
\colhead{$r_{\mathrm{G}}$} \\
\colhead{} &
\colhead{(hh mm ss.s)} &
\colhead{(dd mm ss.s)} &
\colhead{(mJy)} &
\colhead{(mJy)} &
\colhead{(mJy)} &
\colhead{(mJy)} &
\colhead{(pc)} &
\colhead{(kpc)}
}
\startdata
\cutinhead{Star-Forming Regions}
NGC\,0337\,a & 00 59 50.1 & $-$07 34 33.9 & 1.72 $\pm$ 0.13 & 1.25 $\pm$ 0.12 & 0.60 $\pm$ 0.11 & 0.60 $\pm$ 0.11 & 1780 & 0.94 \\
NGC\,0337\,b & 00 59 50.6 & $-$07 34 57.6 & 4.66 $\pm$ 0.13 & 2.53 $\pm$ 0.13 & 1.48 $\pm$ 0.11 & 1.53 $\pm$ 0.11 & 1780 & 1.97 \\
NGC\,0337\,c & 00 59 51.9 & $-$07 34 54.9 & 0.72 $\pm$ 0.14 & 0.52 $\pm$ 0.13 & 0.30 $\pm$ 0.14 & 0.18 $\pm$ 0.11 & 1780 & 3.23 \\
NGC\,0337\,d & 00 59 52.1 & $-$07 34 38.2 & 0.82 $\pm$ 0.14 & 0.83 $\pm$ 0.13 & 0.49 $\pm$ 0.14 & 0.33 $\pm$ 0.11 & 1780 & 4.06 \\
NGC\,0628\,E.\,4 & 01 36 35.7 & +15 50 07.2 & 0.42 $\pm$ 0.10 & 0.33 $\pm$ 0.08 & 0.26 $\pm$ 0.08 & $-$0.09 $\pm$ 0.12 & 800 & 7.61 \\
NGC\,0628\,E.\,2 & 01 36 37.6 & +15 45 07.2 & 0.65 $\pm$ 0.09 & 0.27 $\pm$ 0.11 & 0.49 $\pm$ 0.11 & 0.29 $\pm$ 0.10 & 800 & 4.47 \\
NGC\,0628\,E.\,3 & 01 36 38.7 & +15 44 23.2 & 0.78 $\pm$ 0.09 & 0.72 $\pm$ 0.11 & 0.51 $\pm$ 0.12 & 0.22 $\pm$ 0.12 & 800 & 5.72 \\
NGC\,0628\, & 01 36 41.7 & +15 46 59.0 & $-$0.02 $\pm$ 0.09 & 0.07 $\pm$ 0.15 & 0.05 $\pm$ 0.09 & $-$0.02 $\pm$ 0.08 & 800 & 0.07 \\
NGC\,0628\,E.\,1 & 01 36 45.2 & +15 47 48.3 & 1.17 $\pm$ 0.09 & 0.97 $\pm$ 0.15 & 0.48 $\pm$ 0.11 & 0.23 $\pm$ 0.09 & 800 & 2.48 \\
NGC\,0925\, & 02 27 17.0 & +33 34 43.0 & 0.31 $\pm$ 0.09 & 0.31 $\pm$ 0.09 & 0.18 $\pm$ 0.08 & 0.36 $\pm$ 0.21 & 940 & 0.18 \\
NGC\,2146\,a & 06 18 43.4 & +78 21 08.1 & 50.62 $\pm$ 0.82 & \nodata & 8.51 $\pm$ 0.40 & 6.16 $\pm$ 0.12 & 1580 & 2.08 \\
NGC\,2146\,b & 06 18 38.3 & +78 21 21.4 & 252.64 $\pm$ 0.82 & \nodata & 59.12 $\pm$ 0.35 & 33.39 $\pm$ 0.12 & 1580 & 0.41 \\
NGC\,2146\,c & 06 18 33.7 & +78 21 36.2 & 93.39 $\pm$ 0.82 & \nodata & 22.01 $\pm$ 0.36 & 8.45 $\pm$ 0.12 & 1580 & 1.39 \\
NGC\,2798\, & 09 17 22.8 & +42 00 00.4 & 37.17 $\pm$ 0.09 & 13.02 $\pm$ 0.18 & 4.54 $\pm$ 0.12 & 4.70 $\pm$ 0.09 & 2400 & 0.14 \\
NGC\,2841\, & 09 22 02.6 & +50 58 35.7 & 1.91 $\pm$ 0.06 & 1.55 $\pm$ 0.10 & 1.05 $\pm$ 0.04 & 2.13 $\pm$ 0.11 & 1260 & 0.15 \\
NGC\,3049\, & 09 54 49.5 & +09 16 16.1 & 2.99 $\pm$ 0.10 & 1.74 $\pm$ 0.09 & 1.19 $\pm$ 0.08 & 0.61 $\pm$ 0.11 & 2000 & 0.10 \\
NGC\,3190\, & 10 18 05.6 & +21 49 55.9 & 2.60 $\pm$ 0.07 & 1.21 $\pm$ 0.10 & 0.62 $\pm$ 0.08 & 0.45 $\pm$ 0.12 & 1860 & 0.10 \\
NGC\,3184\, & 10 18 16.9 & +41 25 27.0 & 0.85 $\pm$ 0.06 & 0.41 $\pm$ 0.06 & 0.28 $\pm$ 0.04 & 0.85 $\pm$ 0.14 & 1040 & 0.06 \\
NGC\,3198\, & 10 19 54.9 & +45 32 59.3 & 1.37 $\pm$ 0.09 & 0.69 $\pm$ 0.05 & 0.27 $\pm$ 0.07 & 0.70 $\pm$ 0.16 & 1240 & 0.12 \\
NGC\,3351\,a & 10 43 57.6 & +11 42 08.0 & 12.20 $\pm$ 0.20 & 4.66 $\pm$ 0.11 & 2.97 $\pm$ 0.16 & 2.10 $\pm$ 0.11 & 960 & 0.23 \\
NGC\,3351\,b & 10 43 57.8 & +11 42 18.5 & 12.77 $\pm$ 0.20 & 5.07 $\pm$ 0.11 & 3.26 $\pm$ 0.16 & 1.56 $\pm$ 0.11 & 960 & 0.26 \\
NGC\,3521\,E.\,1 & 11 05 46.3 & $-$00 04 08.9 & 0.05 $\pm$ 0.12 & 0.44 $\pm$ 0.15 & 0.21 $\pm$ 0.11 & 0.12 $\pm$ 0.15 & 1000 & 9.93 \\
NGC\,3521\,E.\,3 & 11 05 47.6 & +00 00 33.4 & 0.11 $\pm$ 0.12 & 0.27 $\pm$ 0.07 & $-$0.04 $\pm$ 0.06 & 0.36 $\pm$ 0.18 & 1000 & 9.51 \\
NGC\,3521\, & 11 05 48.9 & $-$00 02 06.0 & $-$0.11 $\pm$ 0.13 & $-$0.10 $\pm$ 0.23 & $-$0.10 $\pm$ 0.10 & 0.41 $\pm$ 0.12 & 1000 & 0.62 \\
NGC\,3521\,E.\,2a & 11 05 49.3 & $-$00 03 24.2 & 0.92 $\pm$ 0.12 & 0.37 $\pm$ 0.11 & 0.13 $\pm$ 0.08 & 0.53 $\pm$ 0.13 & 1000 & 4.30 \\
NGC\,3521\,E.\,2b & 11 05 49.9 & $-$00 03 55.9 & 0.26 $\pm$ 0.13 & 0.19 $\pm$ 0.11 & 0.13 $\pm$ 0.10 & 0.05 $\pm$ 0.14 & 1000 & 6.04 \\
NGC\,3627\, & 11 20 15.0 & +12 59 29.4 & 9.17 $\pm$ 0.19 & 3.51 $\pm$ 0.34 & 2.23 $\pm$ 0.16 & 1.62 $\pm$ 0.67 & 940 & 0.03 \\
NGC\,3627\,E.\,1 & 11 20 16.3 & +12 57 49.2 & 3.01 $\pm$ 0.17 & 2.66 $\pm$ 0.32 & 2.04 $\pm$ 0.10 & 1.53 $\pm$ 0.77 & 940 & 4.71 \\
NGC\,3627\,E.\,2 & 11 20 16.4 & +12 58 43.4 & 12.53 $\pm$ 0.18 & 7.31 $\pm$ 0.50 & 5.32 $\pm$ 0.11 & 5.98 $\pm$ 0.69 & 940 & 2.75 \\
NGC\,3773\, & 11 38 13.2 & +12 06 43.8 & 2.37 $\pm$ 0.06 & 1.30 $\pm$ 0.09 & 1.06 $\pm$ 0.07 & 1.09 $\pm$ 0.11 & 1340 & 0.02 \\
NGC\,3938\,b & 11 52 48.1 & +44 07 05.9 & $-$0.13 $\pm$ 0.08 & 0.03 $\pm$ 0.05 & 0.21 $\pm$ 0.06 & 0.18 $\pm$ 0.14 & 1720 & 1.41 \\
NGC\,3938\,a & 11 52 49.5 & +44 07 14.0 & $-$0.07 $\pm$ 0.08 & $-$0.05 $\pm$ 0.05 & 0.05 $\pm$ 0.06 & 0.07 $\pm$ 0.15 & 1720 & 0.14 \\
NGC\,3938\,E.\,2a & 11 53 00.5 & +44 08 00.0 & 0.22 $\pm$ 0.07 & 0.13 $\pm$ 0.06 & 0.24 $\pm$ 0.08 & 0.12 $\pm$ 0.20 & 1720 & 11.16 \\
NGC\,3938\,E.\,2b & 11 53 00.1 & +44 07 48.3 & 0.35 $\pm$ 0.07 & 0.29 $\pm$ 0.06 & 0.26 $\pm$ 0.09 & $-$0.22 $\pm$ 0.20 & 1720 & 11.05 \\
NGC\,4254\,E.\,2a & 12 18 45.7 & +14 24 10.4 & 0.27 $\pm$ 0.12 & 0.17 $\pm$ 0.19 & 0.42 $\pm$ 0.07 & 0.13 $\pm$ 0.08 & 1500 & 5.34 \\
NGC\,4254\,E.\,2b & 12 18 46.1 & +14 24 18.8 & 0.19 $\pm$ 0.12 & 0.04 $\pm$ 0.18 & 0.36 $\pm$ 0.06 & 0.10 $\pm$ 0.08 & 1500 & 4.70 \\
NGC\,4254\,a & 12 18 48.6 & +14 24 42.5 & 0.76 $\pm$ 0.13 & 0.32 $\pm$ 0.11 & 0.03 $\pm$ 0.06 & 0.40 $\pm$ 0.08 & 1500 & 1.55 \\
NGC\,4254\,b & 12 18 49.6 & +14 24 59.0 & 0.39 $\pm$ 0.12 & 0.79 $\pm$ 0.11 & 0.37 $\pm$ 0.09 & 1.04 $\pm$ 0.08 & 1500 & 0.11 \\
NGC\,4254\,E.\,1b & 12 18 50.9 & +14 24 06.9 & 0.68 $\pm$ 0.12 & 0.25 $\pm$ 0.12 & 0.57 $\pm$ 0.11 & 0.17 $\pm$ 0.08 & 1500 & 3.89 \\
NGC\,4254\,c & 12 18 50.1 & +14 25 11.6 & 0.39 $\pm$ 0.13 & 0.54 $\pm$ 0.11 & 0.04 $\pm$ 0.06 & 0.93 $\pm$ 0.08 & 1500 & 0.96 \\
NGC\,4254\,E.\,1c & 12 18 50.1 & +14 24 18.6 & 1.05 $\pm$ 0.12 & 0.26 $\pm$ 0.12 & 0.39 $\pm$ 0.11 & 0.27 $\pm$ 0.08 & 1500 & 3.12 \\
NGC\,4254\,d & 12 18 51.6 & +14 25 08.5 & 1.05 $\pm$ 0.13 & 0.24 $\pm$ 0.12 & 0.36 $\pm$ 0.13 & 0.68 $\pm$ 0.08 & 1500 & 2.34 \\
NGC\,4254\,e & 12 18 51.8 & +14 24 49.6 & 0.26 $\pm$ 0.13 & 0.12 $\pm$ 0.12 & 0.63 $\pm$ 0.20 & 0.35 $\pm$ 0.08 & 1500 & 2.81 \\
NGC\,4254\,f & 12 18 51.9 & +14 24 40.9 & 0.17 $\pm$ 0.13 & $-$0.18 $\pm$ 0.13 & 0.33 $\pm$ 0.18 & 0.19 $\pm$ 0.08 & 1500 & 3.15 \\
NGC\,4321\,E.\,2a & 12 22 48.8 & +15 50 12.8 & 0.02 $\pm$ 0.17 & 0.35 $\pm$ 0.14 & 0.11 $\pm$ 0.06 & $-$0.03 $\pm$ 0.17 & 1600 & 8.30 \\
NGC\,4321\,E.\,2b & 12 22 49.9 & +15 50 27.8 & 0.30 $\pm$ 0.16 & 0.14 $\pm$ 0.13 & 0.12 $\pm$ 0.06 & 0.08 $\pm$ 0.17 & 1600 & 7.98 \\
NGC\,4321\,E.\,2 & 12 22 50.6 & +15 50 27.2 & 0.13 $\pm$ 0.16 & $-$0.14 $\pm$ 0.14 & 0.12 $\pm$ 0.06 & 0.05 $\pm$ 0.17 & 1600 & 7.29 \\
NGC\,4321\,a & 12 22 54.6 & +15 49 19.8 & 13.62 $\pm$ 0.16 & 5.51 $\pm$ 0.20 & 3.11 $\pm$ 0.15 & 2.55 $\pm$ 0.15 & 1600 & 0.28 \\
NGC\,4321\,b & 12 22 55.1 & +15 49 20.4 & 16.08 $\pm$ 0.16 & 6.15 $\pm$ 0.20 & 3.52 $\pm$ 0.15 & 3.07 $\pm$ 0.15 & 1600 & 0.27 \\
NGC\,4321\,E.\,1 & 12 22 58.9 & +15 49 35.3 & 0.20 $\pm$ 0.15 & $-$0.12 $\pm$ 0.16 & 0.03 $\pm$ 0.06 & 0.22 $\pm$ 0.15 & 1600 & 4.52 \\
NGC\,4536\, & 12 34 27.6 & +02 11 18.2 & 67.98 $\pm$ 0.15 & 23.36 $\pm$ 0.08 & 14.31 $\pm$ 0.12 & 10.90 $\pm$ 0.14 & 1320 & 0.13 \\
NGC\,4559\,a & 12 35 56.2 & +27 57 40.5 & 0.13 $\pm$ 0.08 & 0.28 $\pm$ 0.10 & 0.16 $\pm$ 0.04 & 0.38 $\pm$ 0.15 & 720 & 1.30 \\
NGC\,4559\,b & 12 35 56.4 & +27 57 21.3 & 0.13 $\pm$ 0.08 & 0.34 $\pm$ 0.10 & 0.13 $\pm$ 0.04 & 0.44 $\pm$ 0.15 & 720 & 1.88 \\
NGC\,4559\,c & 12 35 58.4 & +27 57 29.7 & 0.30 $\pm$ 0.08 & 0.29 $\pm$ 0.10 & 0.23 $\pm$ 0.04 & 0.51 $\pm$ 0.15 & 720 & 0.61 \\
NGC\,4569\, & 12 36 49.8 & +13 09 46.6 & 7.43 $\pm$ 0.14 & 2.79 $\pm$ 0.33 & 1.40 $\pm$ 0.13 & 1.61 $\pm$ 0.15 & 1040 & 0.04 \\
NGC\,4579\, & 12 37 43.5 & +11 49 05.6 & 42.47 $\pm$ 0.31 & 49.48 $\pm$ 0.36 & 57.89 $\pm$ 0.38 & 102.17 $\pm$ 0.12 & 1800 & 0.10 \\
NGC\,4594\,a & 12 39 59.4 & $-$11 37 23.0 & 73.43 $\pm$ 0.12 & 85.41 $\pm$ 0.08 & 68.87 $\pm$ 0.11 & 102.14 $\pm$ 0.09 & 900 & 0.05 \\
NGC\,4625\, & 12 41 52.4 & +41 16 24.0 & $-$0.06 $\pm$ 0.07 & 0.29 $\pm$ 0.06 & 0.11 $\pm$ 0.03 & $-$0.52 $\pm$ 0.59 & 860 & 0.14 \\
NGC\,4631\,a & 12 42 03.4 & +32 32 17.1 & 8.72 $\pm$ 0.11 & 4.16 $\pm$ 0.20 & 2.65 $\pm$ 0.17 & 2.72 $\pm$ 0.11 & 760 & 3.19 \\
NGC\,4631\,b & 12 42 03.9 & +32 32 15.9 & 9.89 $\pm$ 0.11 & 5.55 $\pm$ 0.20 & 3.60 $\pm$ 0.16 & 3.21 $\pm$ 0.11 & 760 & 3.44 \\
NGC\,4631\,c & 12 42 04.3 & +32 32 25.2 & 8.42 $\pm$ 0.11 & 5.00 $\pm$ 0.20 & 4.39 $\pm$ 0.15 & 4.32 $\pm$ 0.11 & 760 & 1.73 \\
NGC\,4631\,d & 12 42 05.9 & +32 32 10.6 & 1.59 $\pm$ 0.11 & 1.32 $\pm$ 0.19 & 0.99 $\pm$ 0.14 & 1.58 $\pm$ 0.11 & 760 & 4.99 \\
NGC\,4631\,e & 12 42 05.5 & +32 32 29.5 & 3.50 $\pm$ 0.11 & 1.59 $\pm$ 0.19 & 2.25 $\pm$ 0.13 & 2.64 $\pm$ 0.11 & 760 & 1.39 \\
NGC\,4631\,E.\,2b & 12 42 21.9 & +32 32 45.0 & 1.47 $\pm$ 0.11 & 1.24 $\pm$ 0.11 & 1.10 $\pm$ 0.08 & 0.78 $\pm$ 0.16 & 760 & 6.65 \\
NGC\,4725\,a & 12 50 26.5 & +25 30 03.0 & 0.13 $\pm$ 0.08 & 0.25 $\pm$ 0.06 & 0.30 $\pm$ 0.03 & 0.25 $\pm$ 0.11 & 1140 & 0.04 \\
NGC\,5194\,E.\,6a & 13 29 39.3 & +47 08 40.7 & 0.78 $\pm$ 0.17 & 0.66 $\pm$ 0.07 & 0.62 $\pm$ 0.04 & 0.06 $\pm$ 0.34 & 700 & 12.32 \\
NGC\,5194\,E.\,3 & 13 29 45.1 & +47 09 57.4 & 0.85 $\pm$ 0.19 & 0.75 $\pm$ 0.12 & 0.74 $\pm$ 0.06 & 0.34 $\pm$ 0.23 & 700 & 7.05 \\
NGC\,5194\,E.\,11a & 13 29 47.1 & +47 13 41.2 & 0.81 $\pm$ 0.21 & 0.21 $\pm$ 0.17 & 0.24 $\pm$ 0.08 & 0.00 $\pm$ 0.22 & 700 & 4.94 \\
NGC\,5194\,E.\,11b & 13 29 47.5 & +47 13 24.7 & 0.19 $\pm$ 0.21 & 0.52 $\pm$ 0.17 & 0.16 $\pm$ 0.06 & 0.49 $\pm$ 0.21 & 700 & 4.34 \\
NGC\,5194\,E.\,11d & 13 29 49.5 & +47 13 28.7 & 0.02 $\pm$ 0.21 & 0.07 $\pm$ 0.17 & 0.12 $\pm$ 0.05 & $-$0.11 $\pm$ 0.21 & 700 & 4.08 \\
NGC\,5194\,E.\,11c & 13 29 49.6 & +47 14 00.2 & 1.27 $\pm$ 0.21 & 0.53 $\pm$ 0.18 & 0.13 $\pm$ 0.08 & 0.00 $\pm$ 0.23 & 700 & 5.22 \\
NGC\,5194\,c & 13 29 50.2 & +47 11 31.8 & 2.95 $\pm$ 0.25 & 1.51 $\pm$ 0.21 & 0.41 $\pm$ 0.13 & 0.70 $\pm$ 0.18 & 700 & 1.80 \\
NGC\,5194\,E.\,11e & 13 29 50.6 & +47 13 44.9 & 0.65 $\pm$ 0.21 & 0.62 $\pm$ 0.17 & 0.36 $\pm$ 0.06 & $-$0.03 $\pm$ 0.22 & 700 & 4.63 \\
NGC\,5194\,b & 13 29 51.6 & +47 12 06.7 & 3.77 $\pm$ 0.24 & 1.70 $\pm$ 0.20 & 0.63 $\pm$ 0.12 & 0.35 $\pm$ 0.18 & 700 & 0.98 \\
NGC\,5194\,e & 13 29 52.5 & +47 11 52.6 & 13.64 $\pm$ 0.24 & 4.80 $\pm$ 0.20 & 1.80 $\pm$ 0.10 & 1.21 $\pm$ 0.18 & 700 & 0.36 \\
NGC\,5194\,d & 13 29 52.7 & +47 11 40.6 & 15.17 $\pm$ 0.24 & 4.66 $\pm$ 0.20 & 1.57 $\pm$ 0.10 & 1.58 $\pm$ 0.18 & 700 & 0.09 \\
NGC\,5194\,E.\,1c & 13 29 53.1 & +47 12 39.4 & 0.86 $\pm$ 0.22 & 0.22 $\pm$ 0.15 & 0.47 $\pm$ 0.10 & 0.32 $\pm$ 0.19 & 700 & 2.32 \\
NGC\,5194\,E.\,4a & 13 29 53.9 & +47 14 04.8 & $-$0.08 $\pm$ 0.18 & 0.04 $\pm$ 0.19 & 0.19 $\pm$ 0.05 & 0.43 $\pm$ 0.24 & 700 & 5.88 \\
NGC\,5194\,E.\,10a & 13 29 55.3 & +47 10 47.1 & 0.60 $\pm$ 0.23 & 0.34 $\pm$ 0.15 & 0.26 $\pm$ 0.10 & $-$0.13 $\pm$ 0.18 & 700 & 2.34 \\
NGC\,5194\,E.\,4c & 13 29 55.6 & +47 13 50.2 & 0.76 $\pm$ 0.18 & 0.16 $\pm$ 0.15 & 0.29 $\pm$ 0.05 & 0.45 $\pm$ 0.24 & 700 & 5.80 \\
NGC\,5194\,a & 13 29 55.7 & +47 11 45.9 & 1.62 $\pm$ 0.24 & 1.12 $\pm$ 0.21 & 1.03 $\pm$ 0.15 & 0.78 $\pm$ 0.18 & 700 & 1.91 \\
NGC\,5194\,E.\,10b & 13 29 56.5 & +47 10 46.9 & 0.95 $\pm$ 0.23 & 0.55 $\pm$ 0.14 & 0.47 $\pm$ 0.09 & 0.01 $\pm$ 0.18 & 700 & 2.72 \\
NGC\,5194\,E.\,4d & 13 29 58.7 & +47 14 09.3 & 0.66 $\pm$ 0.18 & 0.28 $\pm$ 0.14 & 0.21 $\pm$ 0.05 & 0.51 $\pm$ 0.28 & 700 & 7.70 \\
NGC\,5194\,E.\,5a & 13 29 59.6 & +47 13 59.8 & 0.55 $\pm$ 0.18 & 0.47 $\pm$ 0.19 & 0.48 $\pm$ 0.08 & 0.69 $\pm$ 0.27 & 700 & 7.74 \\
NGC\,5194\,E.\,9 & 13 29 59.7 & +47 11 12.3 & 0.29 $\pm$ 0.22 & 0.27 $\pm$ 0.11 & 0.42 $\pm$ 0.08 & $-$0.24 $\pm$ 0.19 & 700 & 4.13 \\
NGC\,5194\,E.\,7a & 13 30 01.2 & +47 09 28.5 & 0.30 $\pm$ 0.20 & 0.38 $\pm$ 0.09 & 0.18 $\pm$ 0.07 & $-$0.37 $\pm$ 0.25 & 700 & 6.22 \\
NGC\,5194\,E.\,8 & 13 30 01.4 & +47 12 51.7 & 1.29 $\pm$ 0.20 & 1.06 $\pm$ 0.12 & 0.60 $\pm$ 0.06 & 0.41 $\pm$ 0.21 & 700 & 6.65 \\
NGC\,5194\,E.\,7b & 13 30 02.3 & +47 09 48.7 & 1.65 $\pm$ 0.20 & 0.60 $\pm$ 0.09 & 0.52 $\pm$ 0.05 & $-$0.10 $\pm$ 0.24 & 700 & 6.33 \\
NGC\,5194\,E.\,7c & 13 30 03.4 & +47 09 40.3 & 0.68 $\pm$ 0.20 & 0.14 $\pm$ 0.09 & 0.16 $\pm$ 0.06 & 0.17 $\pm$ 0.25 & 700 & 6.96 \\
NGC\,5474\, & 14 05 01.3 & +53 39 44.0 & 0.11 $\pm$ 0.07 & 0.06 $\pm$ 0.06 & 0.05 $\pm$ 0.03 & $-$0.22 $\pm$ 0.26 & 640 & 0.07 \\
NGC\,5713\,E.\,2a & 14 40 10.8 & $-$00 17 35.5 & 4.68 $\pm$ 0.10 & 3.25 $\pm$ 0.32 & 1.32 $\pm$ 0.12 & 2.07 $\pm$ 0.17 & 2200 & 1.98 \\
NGC\,5713\,E.\,2b & 14 40 10.8 & $-$00 17 50.2 & 1.80 $\pm$ 0.10 & 1.25 $\pm$ 0.33 & 0.19 $\pm$ 0.10 & 1.10 $\pm$ 0.17 & 2200 & 3.29 \\
NGC\,5713\, & 14 40 11.3 & $-$00 17 27.0 & 15.05 $\pm$ 0.10 & 7.68 $\pm$ 0.36 & 4.07 $\pm$ 0.13 & 4.40 $\pm$ 0.17 & 2200 & 0.80 \\
NGC\,5713\,E.\,1 & 14 40 11.3 & $-$00 17 18.2 & 19.81 $\pm$ 0.10 & 9.06 $\pm$ 0.33 & 4.72 $\pm$ 0.11 & 3.95 $\pm$ 0.17 & 2200 & 0.31 \\
NGC\,6946\,E.\,4c & 20 34 22.7 & +60 10 34.1 & 2.58 $\pm$ 0.17 & 2.06 $\pm$ 0.07 & 1.45 $\pm$ 0.07 & 2.15 $\pm$ 0.19 & 640 & 8.74 \\
NGC\,6946\,E.\,8 & 20 34 32.2 & +60 10 19.3 & 3.37 $\pm$ 0.17 & 1.39 $\pm$ 0.06 & 1.42 $\pm$ 0.07 & 1.29 $\pm$ 0.17 & 640 & 6.12 \\
NGC\,6946\,E.\,5a & 20 34 37.1 & +60 05 10.9 & $-$0.15 $\pm$ 0.17 & $-$0.07 $\pm$ 0.06 & 0.14 $\pm$ 0.05 & 0.09 $\pm$ 0.19 & 640 & 9.24 \\
NGC\,6946\,E.\,5b & 20 34 39.3 & +60 04 52.4 & 0.40 $\pm$ 0.17 & 0.48 $\pm$ 0.05 & 0.39 $\pm$ 0.04 & 0.61 $\pm$ 0.19 & 640 & 9.70 \\
NGC\,6946\,E.\,3a & 20 34 49.8 & +60 12 40.6 & 0.06 $\pm$ 0.15 & 0.42 $\pm$ 0.06 & 0.28 $\pm$ 0.05 & 0.58 $\pm$ 0.18 & 640 & 7.74 \\
NGC\,6946\,E.\,3b & 20 34 52.2 & +60 12 43.7 & 0.54 $\pm$ 0.15 & 0.72 $\pm$ 0.06 & 0.59 $\pm$ 0.04 & 0.99 $\pm$ 0.18 & 640 & 7.73 \\
NGC\,6946\,b & 20 34 52.2 & +60 09 14.3 & 48.66 $\pm$ 0.17 & 21.14 $\pm$ 0.12 & 12.91 $\pm$ 0.22 & 12.30 $\pm$ 0.12 & 640 & 0.02 \\
NGC\,6946\,E.\,6b & 20 35 06.9 & +60 10 46.5 & 0.83 $\pm$ 0.15 & 0.69 $\pm$ 0.08 & 0.42 $\pm$ 0.10 & 0.69 $\pm$ 0.18 & 640 & 4.73 \\
NGC\,6946\,E.\,9 & 20 35 11.8 & +60 08 57.4 & 1.96 $\pm$ 0.15 & 1.96 $\pm$ 0.07 & 1.67 $\pm$ 0.09 & 2.23 $\pm$ 0.16 & 640 & 5.07 \\
NGC\,6946\,E.\,7 & 20 35 12.9 & +60 08 50.5 & 0.68 $\pm$ 0.15 & 1.48 $\pm$ 0.07 & 0.92 $\pm$ 0.09 & 1.18 $\pm$ 0.17 & 640 & 5.64 \\
NGC\,6946\,E.\,1 & 20 35 16.8 & +60 11 00.0 & 0.91 $\pm$ 0.15 & 0.62 $\pm$ 0.05 & 0.68 $\pm$ 0.07 & 0.35 $\pm$ 0.19 & 640 & 6.99 \\
NGC\,7331\, & 22 37 04.1 & +34 24 56.0 & $-$0.07 $\pm$ 0.15 & 1.18 $\pm$ 0.51 & $-$0.37 $\pm$ 0.15 & 0.52 $\pm$ 0.11 & 1340 & 0.00 \\
\cutinhead{Likely Associated with Supernovae}
NGC\,6946\,E.\,6a & 20 35 06.8 & +60 10 58.5 & 2.07 $\pm$ 0.15 & 1.18 $\pm$ 0.08 & 1.40 $\pm$ 0.10 & 1.67 $\pm$ 0.18 & 640 & 4.86 \\
\cutinhead{Likely AME Candidates}
NGC\,4254\,E.\,1a & 12 18 49.2 & +14 23 57.9 & 0.29 $\pm$ 0.12 & 0.14 $\pm$ 0.12 & 0.38 $\pm$ 0.09 & 0.15 $\pm$ 0.08 & 1500 & 4.43 \\
NGC\,4725\,b & 12 50 28.4 & +25 30 21.8 & 0.18 $\pm$ 0.09 & 0.20 $\pm$ 0.07 & 0.35 $\pm$ 0.05 & 0.25 $\pm$ 0.11 & 1140 & 1.92 \\
NGC\,5194\,E.\,2 & 13 29 44.1 & +47 10 23.4 & 1.58 $\pm$ 0.19 & 0.81 $\pm$ 0.09 & 0.90 $\pm$ 0.07 & 0.54 $\pm$ 0.21 & 700 & 6.83 \\
NGC\,5194\,E.\,1a & 13 29 49.5 & +47 12 40.2 & 1.01 $\pm$ 0.22 & 0.91 $\pm$ 0.17 & 1.13 $\pm$ 0.17 & 0.35 $\pm$ 0.19 & 700 & 2.53 \\
NGC\,5194\,E.\,1b & 13 29 52.7 & +47 12 43.6 & 0.68 $\pm$ 0.22 & 0.65 $\pm$ 0.15 & 0.70 $\pm$ 0.10 & 0.44 $\pm$ 0.19 & 700 & 2.32 \\
NGC\,5194\,E.\,4b & 13 29 55.4 & +47 14 01.6 & 0.90 $\pm$ 0.18 & 0.26 $\pm$ 0.15 & 0.49 $\pm$ 0.04 & 0.32 $\pm$ 0.25 & 700 & 6.19 \\
NGC\,6946\,E.\,4a & 20 34 19.8 & +60 10 06.6 & 1.96 $\pm$ 0.18 & 1.90 $\pm$ 0.06 & 1.98 $\pm$ 0.04 & 1.45 $\pm$ 0.19 & 640 & 9.04 \\
NGC\,6946\,E.\,4b & 20 34 21.4 & +60 10 17.6 & 0.17 $\pm$ 0.18 & 1.22 $\pm$ 0.07 & 0.73 $\pm$ 0.05 & 1.34 $\pm$ 0.19 & 640 & 8.80 \\
NGC\,6946\,a & 20 34 51.2 & +60 09 39.2 & 1.09 $\pm$ 0.17 & 1.37 $\pm$ 0.13 & 1.22 $\pm$ 0.29 & 1.50 $\pm$ 0.13 & 640 & 1.00 \\
NGC\,6946\,c & 20 34 52.7 & +60 09 30.5 & 3.76 $\pm$ 0.17 & 2.89 $\pm$ 0.12 & 1.31 $\pm$ 0.23 & 1.81 $\pm$ 0.12 & 640 & 0.60 \\
NGC\,6946\,E.\,2a & 20 35 23.5 & +60 09 48.8 & 0.18 $\pm$ 0.15 & 0.69 $\pm$ 0.16 & 0.34 $\pm$ 0.08 & 1.22 $\pm$ 0.19 & 640 & 8.12 \\
NGC\,6946\,E.\,2b & 20 35 25.3 & +60 09 58.8 & 2.21 $\pm$ 0.16 & 2.18 $\pm$ 0.16 & 1.76 $\pm$ 0.08 & 2.01 $\pm$ 0.19 & 640 & 8.60 \\
\enddata
\tablecomments{We report all flux density values as measured even if they are non-detections (S/N $<$ 3). The standard flux calibration uncertainties for the VLA and GBT ($\sim3\%$ and $\sim10\%$, respectively) are not included in the photometric uncertainties shown here.}
\end{deluxetable*}

The VLA photometric uncertainties for each of the three corresponding frequencies (3, 15, and 33\,GHz) were estimated by taking the scaled median absolute deviation of the non-primary beam corrected images ($\mathrm{MADN}_{\nu}$) and multiplying by the square root of the number of beams ($\mathrm{n_{bm}}$) within each photometric aperture, as well as the primary beam correction factor, defined as the ratio of the photometric flux density measured in the primary beam corrected maps ($f_{\nu,\mathrm{corr}}$) to the photometric flux density measured in the non-primary beam corrected maps ($f_{\nu}$):

\begin{equation} \label{vla_errors}
\mathrm{\sigma_{\,VLA}}=\sqrt{\mathrm{n_{bms}}}\ \mathrm{MADN}_{\nu} \left( \frac{f_{\nu,\mathrm{corr}}}{f_{\nu}} \right).
\end{equation}

The GBT (90\,GHz) photometric uncertainties were estimated using the following process:

\begin{enumerate}
    \item Create a mask from the weight map by selecting values greater than 0.2 times the median of the non-zero values in the weight map. The weight map itself is created by calculating the standard deviation for each detector and gridding the weights (defined as the inverse of the standard deviation) into the map.
    \item Create a per-pixel signal-to-noise ratio map by multiplying the data map and the square root of the weight map, applying the mask from the previous step.
    \item Calculate the correction factor ($\kappa$) by taking the square of the scaled scatter of the map created in Step 2. It is expected that most pixels in our signal-to-noise ratio map have no signal, resulting in a standard deviation of unity; however, because the noise estimate is imperfect, it is necessary to calculate and apply a statistical correction to the data. We found an average $\kappa$ of $\approx$1.7, with a scatter of $\approx$0.5.
    \item Calculate 
    \begin{equation}
        \gamma=\left[ \frac{\mathrm{p_{scl}}}{\left(\frac{\pi}{4\ln{2}} \right) \theta_{1/2}} \right] ^{2}
    \end{equation}
    where $\gamma$ is the factor for converting from an aperture sum to a flux density, $\mathrm{p_{scl}}$ is the pixel scale, and $\theta_{1/2}$ is the FWHM of the beam.
    \item Calculate the median value of the weight map within the appropriate aperture ($\mathrm{med_{ap}}$).
    \item Calculate the number of pixels within the appropriate aperture ($\mathrm{n_{pix}}$).
    \item Use the following expression to calculate the final photometric uncertainty:
        \begin{equation}
            \mathrm{\sigma_{\,GBT}}=\gamma\ \sqrt{\frac{\mathrm{n_{pix}}\kappa}{\mathrm{med_{ap}}}}\end{equation}
    
\end{enumerate}

\section{RESULTS}
\label{section:RESULTS}

We present spectral index and thermal fraction distributions calculated using a basic single and two-component power law fit and more complex Monte Carlo Markov Chain (MCMC) fitting (Table \ref{tab:parameters}). For single and two-component power law fitting, we only fit regions in our sample with a S/N $>$ 3 measured in at least two radio bands. This criterion removed 32 regions, reducing the sample used to calculate the spectral indices from 3 to 33\,GHz and thermal fractions at 33\, GHz for power law fitting to 87 sources. 
The median signal-to-noise of the remaining 87 sources is 13, 9, 10, and 5 for detections at 3, 15, 33, and 90\,GHz, respectively. 
For these regions, we did not fit the non-detections for single or two-component power law fitting; however, we used all 119 regions in our sample and both detections and non-detections for MCMC fitting. 

\startlongtable
\begin{deluxetable*}{l|cc|cc|cc} \label{tab:parameters}
\tablecolumns{7}
\tablecaption{Derived Parameters at 10\arcsec Angular Resolution}
\tablehead{
\colhead{Source ID} &
\multicolumn{2}{c}{3\textendash33\,GHz power law} &
\multicolumn{2}{c}{MCMC w/ dust \& 90\,GHz} &
\multicolumn{2}{c}{MCMC w/o dust \& 90\,GHz}\\
\colhead{} &
\colhead{$\rm{\alpha^{3\textendash33\,GHz}}$} &
\colhead{$f_{\mathrm{T}}^{\mathrm{33\,GHz}}$} &
\colhead{$\rm{\alpha^{NT}}$} &
\colhead{$f_{\mathrm{T}}^{\mathrm{33\,GHz}}$} &
\colhead{$\rm{\alpha^{NT}}$} &
\colhead{$f_{\mathrm{T}}^{\mathrm{33\,GHz}}$}
}
\startdata
\cutinhead{Star-Forming Regions}
NGC\,0337\,a & $-$0.31 $\pm$ 0.06 & 0.86 $\pm$ 0.05 & ${-0.72}^{0.13}_{-0.13}$ & 0.75 $\pm$ 0.13 & ${-0.74}^{0.13}_{-0.13}$ & 0.80 $\pm$ 0.11 \\
NGC\,0337\,b & $-$0.43 $\pm$ 0.03 & 0.75 $\pm$ 0.03 & ${-0.67}^{0.12}_{-0.12}$ & 0.59 $\pm$ 0.18 & ${-0.67}^{0.12}_{-0.12}$ & 0.59 $\pm$ 0.18 \\
NGC\,0337\,c & $-$0.21 $\pm$ 0.19 & 0.94 $\pm$ 0.13 & ${-0.77}^{0.13}_{-0.13}$ & 0.72 $\pm$ 0.20 & ${-0.78}^{0.13}_{-0.13}$ & 0.86 $\pm$ 0.12 \\
NGC\,0337\,d & $-$0.10 $\pm$ 0.12 & 1.00 $\pm$ 0.06 & ${-0.76}^{0.14}_{-0.14}$ & 0.90 $\pm$ 0.08 & ${-0.77}^{0.13}_{-0.13}$ & 0.95 $\pm$ 0.05 \\
NGC\,0628\,Enuc.\,4 & $-$0.19 $\pm$ 0.15 & 0.95 $\pm$ 0.10 & ${-0.77}^{0.13}_{-0.13}$ & 0.80 $\pm$ 0.17 & ${-0.78}^{0.13}_{-0.13}$ & 0.91 $\pm$ 0.09 \\
NGC\,0628\,Enuc.\,2 & $-$0.11 $\pm$ 0.11 & 0.99 $\pm$ 0.06 & ${-0.78}^{0.13}_{-0.13}$ & 0.84 $\pm$ 0.12 & ${-0.78}^{0.13}_{-0.13}$ & 0.90 $\pm$ 0.09 \\
NGC\,0628\,Enuc.\,3 & $-$0.13 $\pm$ 0.09 & 0.99 $\pm$ 0.05 & ${-0.76}^{0.13}_{-0.13}$ & 0.90 $\pm$ 0.08 & ${-0.78}^{0.13}_{-0.13}$ & 0.96 $\pm$ 0.04 \\
NGC\,0628\, & \nodata & \nodata & ${-0.80}^{0.13}_{-0.13}$ & 0.72 $\pm$ 0.26 & ${-0.79}^{0.13}_{-0.13}$ & 0.84 $\pm$ 0.18 \\
NGC\,0628\,Enuc.\,1 & $-$0.25 $\pm$ 0.08 & 0.91 $\pm$ 0.06 & ${-0.72}^{0.12}_{-0.13}$ & 0.67 $\pm$ 0.16 & ${-0.77}^{0.13}_{-0.13}$ & 0.87 $\pm$ 0.08 \\
NGC\,0925\, & $-$0.01 $\pm$ 0.25 & 1.00 $\pm$ 0.10 & ${-0.78}^{0.13}_{-0.13}$ & 0.86 $\pm$ 0.12 & ${-0.78}^{0.13}_{-0.13}$ & 0.91 $\pm$ 0.09 \\
NGC\,2146\,a & $-$0.74 $\pm$ 0.02 & 0.23 $\pm$ 0.05 & ${-0.85}^{0.07}_{-0.10}$ & 0.27 $\pm$ 0.19 & ${-0.85}^{0.07}_{-0.11}$ & 0.27 $\pm$ 0.19 \\
NGC\,2146\,b & $-$0.61 $\pm$ 0.00 & 0.50 $\pm$ 0.00 & ${-0.63}^{0.01}_{-0.02}$ & 0.07 $\pm$ 0.05 & ${-0.80}^{0.12}_{-0.12}$ & 0.46 $\pm$ 0.20 \\
NGC\,2146\,c & $-$0.60 $\pm$ 0.01 & 0.51 $\pm$ 0.01 & ${-0.69}^{0.00}_{-0.01}$ & 0.00 $\pm$ 0.00 & ${-0.79}^{0.11}_{-0.12}$ & 0.46 $\pm$ 0.20 \\
NGC\,2798\, & $-$0.74 $\pm$ 0.01 & 0.24 $\pm$ 0.02 & ${-0.78}^{0.01}_{-0.01}$ & 0.01 $\pm$ 0.01 & ${-0.78}^{0.01}_{-0.01}$ & 0.01 $\pm$ 0.01 \\
NGC\,2841\, & $-$0.24 $\pm$ 0.02 & 0.91 $\pm$ 0.01 & ${-0.74}^{0.13}_{-0.13}$ & 0.85 $\pm$ 0.05 & ${-0.74}^{0.13}_{-0.13}$ & 0.90 $\pm$ 0.05 \\
NGC\,3049\, & $-$0.37 $\pm$ 0.03 & 0.81 $\pm$ 0.03 & ${-0.58}^{0.11}_{-0.13}$ & 0.52 $\pm$ 0.21 & ${-0.72}^{0.13}_{-0.13}$ & 0.75 $\pm$ 0.11 \\
NGC\,3190\, & $-$0.54 $\pm$ 0.04 & 0.60 $\pm$ 0.06 & ${-0.71}^{0.09}_{-0.11}$ & 0.38 $\pm$ 0.21 & ${-0.73}^{0.10}_{-0.12}$ & 0.46 $\pm$ 0.21 \\
NGC\,3184\, & $-$0.46 $\pm$ 0.06 & 0.72 $\pm$ 0.07 & ${-0.78}^{0.13}_{-0.13}$ & 0.61 $\pm$ 0.16 & ${-0.77}^{0.13}_{-0.13}$ & 0.69 $\pm$ 0.15 \\
NGC\,3198\, & $-$0.47 $\pm$ 0.06 & 0.69 $\pm$ 0.07 & ${-0.72}^{0.11}_{-0.12}$ & 0.52 $\pm$ 0.21 & ${-0.72}^{0.11}_{-0.12}$ & 0.54 $\pm$ 0.21 \\
NGC\,3351\,a & $-$0.59 $\pm$ 0.02 & 0.52 $\pm$ 0.03 & ${-0.70}^{0.07}_{-0.07}$ & 0.33 $\pm$ 0.17 & ${-0.73}^{0.08}_{-0.10}$ & 0.39 $\pm$ 0.19 \\
NGC\,3351\,b & $-$0.57 $\pm$ 0.01 & 0.56 $\pm$ 0.02 & ${-0.61}^{0.02}_{-0.03}$ & 0.06 $\pm$ 0.05 & ${-0.71}^{0.08}_{-0.09}$ & 0.39 $\pm$ 0.18 \\
NGC\,3521\,Enuc.\,1 & \nodata & \nodata & ${-0.78}^{0.13}_{-0.13}$ & 0.88 $\pm$ 0.11 & ${-0.78}^{0.13}_{-0.13}$ & 0.94 $\pm$ 0.07 \\
NGC\,3521\,Enuc.\,3 & \nodata & \nodata & ${-0.79}^{0.12}_{-0.13}$ & 0.65 $\pm$ 0.26 & ${-0.79}^{0.12}_{-0.13}$ & 0.75 $\pm$ 0.25 \\
NGC\,3521\, & \nodata & \nodata & ${-0.80}^{0.13}_{-0.13}$ & 0.49 $\pm$ 0.29 & ${-0.80}^{0.13}_{-0.13}$ & 0.76 $\pm$ 0.25 \\
NGC\,3521\,Enuc.\,2a & $-$0.58 $\pm$ 0.21 & 0.55 $\pm$ 0.33 & ${-0.82}^{0.11}_{-0.12}$ & 0.39 $\pm$ 0.25 & ${-0.82}^{0.11}_{-0.12}$ & 0.42 $\pm$ 0.26 \\
NGC\,3521\,Enuc.\,2b & \nodata & \nodata & ${-0.79}^{0.13}_{-0.13}$ & 0.73 $\pm$ 0.24 & ${-0.79}^{0.13}_{-0.13}$ & 0.81 $\pm$ 0.19 \\
NGC\,3627\, & $-$0.59 $\pm$ 0.03 & 0.53 $\pm$ 0.05 & ${-0.77}^{0.10}_{-0.12}$ & 0.42 $\pm$ 0.20 & ${-0.78}^{0.10}_{-0.12}$ & 0.47 $\pm$ 0.19 \\
NGC\,3627\,Enuc.\,1 & $-$0.16 $\pm$ 0.03 & 0.97 $\pm$ 0.02 & ${-0.78}^{0.14}_{-0.13}$ & 0.96 $\pm$ 0.03 & ${-0.78}^{0.13}_{-0.13}$ & 0.97 $\pm$ 0.02 \\
NGC\,3627\,Enuc.\,2 & $-$0.36 $\pm$ 0.01 & 0.82 $\pm$ 0.01 & ${-0.74}^{0.13}_{-0.13}$ & 0.76 $\pm$ 0.09 & ${-0.75}^{0.13}_{-0.13}$ & 0.78 $\pm$ 0.09 \\
NGC\,3773\, & $-$0.35 $\pm$ 0.02 & 0.83 $\pm$ 0.02 & ${-0.78}^{0.12}_{-0.12}$ & 0.81 $\pm$ 0.08 & ${-0.78}^{0.13}_{-0.13}$ & 0.82 $\pm$ 0.07 \\
NGC\,3938\,b & \nodata & \nodata & ${-0.78}^{0.13}_{-0.13}$ & 0.73 $\pm$ 0.20 & ${-0.79}^{0.13}_{-0.13}$ & 0.90 $\pm$ 0.11 \\
NGC\,3938\,a & \nodata & \nodata & ${-0.80}^{0.13}_{-0.13}$ & 0.55 $\pm$ 0.30 & ${-0.80}^{0.13}_{-0.13}$ & 0.77 $\pm$ 0.24 \\
NGC\,3938\,Enuc.\,2a & \nodata & \nodata & ${-0.79}^{0.13}_{-0.13}$ & 0.84 $\pm$ 0.14 & ${-0.78}^{0.13}_{-0.13}$ & 0.90 $\pm$ 0.11 \\
NGC\,3938\,Enuc.\,2b & $-$0.11 $\pm$ 0.14 & 0.99 $\pm$ 0.07 & ${-0.77}^{0.13}_{-0.13}$ & 0.90 $\pm$ 0.09 & ${-0.78}^{0.13}_{-0.13}$ & 0.94 $\pm$ 0.06 \\
NGC\,4254\,Enuc.\,2a & \nodata & \nodata & ${-0.78}^{0.13}_{-0.13}$ & 0.95 $\pm$ 0.05 & ${-0.78}^{0.13}_{-0.13}$ & 0.97 $\pm$ 0.03 \\
NGC\,4254\,Enuc.\,2b & \nodata & \nodata & ${-0.78}^{0.13}_{-0.13}$ & 0.94 $\pm$ 0.06 & ${-0.78}^{0.13}_{-0.13}$ & 0.97 $\pm$ 0.04 \\
NGC\,4254\,a & \nodata & \nodata & ${-0.86}^{0.11}_{-0.11}$ & 0.27 $\pm$ 0.22 & ${-0.86}^{0.11}_{-0.11}$ & 0.30 $\pm$ 0.23 \\
NGC\,4254\,b & 0.01 $\pm$ 0.17 & 1.00 $\pm$ 0.07 & ${-0.77}^{0.13}_{-0.13}$ & 0.92 $\pm$ 0.03 & ${-0.78}^{0.13}_{-0.13}$ & 0.98 $\pm$ 0.03 \\
NGC\,4254\,Enuc.\,1b & $-$0.08 $\pm$ 0.11 & 1.00 $\pm$ 0.05 & ${-0.78}^{0.13}_{-0.13}$ & 0.79 $\pm$ 0.15 & ${-0.78}^{0.13}_{-0.13}$ & 0.92 $\pm$ 0.07 \\
NGC\,4254\,c & 0.20 $\pm$ 0.24 & 1.00 $\pm$ 0.06 & ${-0.79}^{0.12}_{-0.12}$ & 0.46 $\pm$ 0.27 & ${-0.78}^{0.12}_{-0.13}$ & 0.61 $\pm$ 0.28 \\
NGC\,4254\,Enuc.\,1c & $-$0.41 $\pm$ 0.13 & 0.77 $\pm$ 0.14 & ${-0.81}^{0.12}_{-0.12}$ & 0.52 $\pm$ 0.24 & ${-0.81}^{0.12}_{-0.12}$ & 0.59 $\pm$ 0.23 \\
NGC\,4254\,d & \nodata & \nodata & ${-0.82}^{0.12}_{-0.12}$ & 0.48 $\pm$ 0.26 & ${-0.82}^{0.12}_{-0.12}$ & 0.54 $\pm$ 0.25 \\
NGC\,4254\,e & \nodata & \nodata & ${-0.78}^{0.13}_{-0.13}$ & 0.85 $\pm$ 0.13 & ${-0.79}^{0.13}_{-0.13}$ & 0.90 $\pm$ 0.12 \\
NGC\,4254\,f & \nodata & \nodata & ${-0.80}^{0.13}_{-0.13}$ & 0.68 $\pm$ 0.27 & ${-0.80}^{0.13}_{-0.13}$ & 0.76 $\pm$ 0.25 \\
NGC\,4321\,Enuc.\,2a & \nodata & \nodata & ${-0.78}^{0.13}_{-0.13}$ & 0.80 $\pm$ 0.19 & ${-0.78}^{0.13}_{-0.13}$ & 0.86 $\pm$ 0.14 \\
NGC\,4321\,Enuc.\,2b & \nodata & \nodata & ${-0.79}^{0.12}_{-0.12}$ & 0.66 $\pm$ 0.27 & ${-0.79}^{0.13}_{-0.13}$ & 0.72 $\pm$ 0.25 \\
NGC\,4321\,Enuc.\,2 & \nodata & \nodata & ${-0.80}^{0.13}_{-0.13}$ & 0.69 $\pm$ 0.26 & ${-0.79}^{0.13}_{-0.13}$ & 0.78 $\pm$ 0.22 \\
NGC\,4321\,a & $-$0.59 $\pm$ 0.02 & 0.53 $\pm$ 0.03 & ${-0.66}^{0.05}_{-0.08}$ & 0.21 $\pm$ 0.17 & ${-0.66}^{0.05}_{-0.08}$ & 0.22 $\pm$ 0.17 \\
NGC\,4321\,b & $-$0.62 $\pm$ 0.01 & 0.48 $\pm$ 0.02 & ${-0.68}^{0.04}_{-0.07}$ & 0.20 $\pm$ 0.16 & ${-0.68}^{0.04}_{-0.07}$ & 0.19 $\pm$ 0.16 \\
NGC\,4321\,Enuc.\,1 & \nodata & \nodata & ${-0.81}^{0.13}_{-0.13}$ & 0.54 $\pm$ 0.31 & ${-0.81}^{0.12}_{-0.13}$ & 0.64 $\pm$ 0.31 \\
NGC\,4536\, & $-$0.66 $\pm$ 0.00 & 0.41 $\pm$ 0.00 & ${-0.71}^{0.02}_{-0.02}$ & 0.17 $\pm$ 0.06 & ${-0.72}^{0.02}_{-0.02}$ & 0.20 $\pm$ 0.05 \\
NGC\,4559\,a & \nodata & \nodata & ${-0.78}^{0.13}_{-0.13}$ & 0.87 $\pm$ 0.08 & ${-0.78}^{0.13}_{-0.13}$ & 0.94 $\pm$ 0.06 \\
NGC\,4559\,b & $-$1.22 $\pm$ 0.55 & 0.23 $\pm$ 1.09 & ${-0.78}^{0.13}_{-0.13}$ & 0.84 $\pm$ 0.10 & ${-0.78}^{0.13}_{-0.13}$ & 0.93 $\pm$ 0.07 \\
NGC\,4559\,c & $-$0.12 $\pm$ 0.14 & 0.99 $\pm$ 0.07 & ${-0.78}^{0.13}_{-0.13}$ & 0.88 $\pm$ 0.07 & ${-0.78}^{0.13}_{-0.13}$ & 0.94 $\pm$ 0.06 \\
NGC\,4569\, & $-$0.68 $\pm$ 0.04 & 0.37 $\pm$ 0.07 & ${-0.80}^{0.08}_{-0.11}$ & 0.29 $\pm$ 0.20 & ${-0.79}^{0.08}_{-0.10}$ & 0.29 $\pm$ 0.20 \\
NGC\,4579\, & 0.13 $\pm$ 0.00 & 1.00 $\pm$ 0.00 & ${0.18}^{0.08}_{-0.04}$ & 0.15 $\pm$ 0.11 & ${0.27}^{0.06}_{-0.06}$ & 0.27 $\pm$ 0.07 \\
NGC\,4594\,a & 0.00 $\pm$ 0.00 & 1.00 $\pm$ 0.00 & ${0.00}^{0.00}_{0.00}$ & 0.01 $\pm$ 0.01 & ${0.00}^{0.00}_{0.00}$ & 0.01 $\pm$ 0.01 \\
NGC\,4625\, & $-$1.22 $\pm$ 0.45 & 0.23 $\pm$ 0.89 & ${-0.78}^{0.13}_{-0.13}$ & 0.87 $\pm$ 0.10 & ${-0.78}^{0.13}_{-0.13}$ & 0.96 $\pm$ 0.04 \\
NGC\,4631\,a & $-$0.48 $\pm$ 0.02 & 0.69 $\pm$ 0.03 & ${-0.70}^{0.11}_{-0.13}$ & 0.54 $\pm$ 0.19 & ${-0.70}^{0.11}_{-0.12}$ & 0.56 $\pm$ 0.18 \\
NGC\,4631\,b & $-$0.40 $\pm$ 0.02 & 0.78 $\pm$ 0.02 & ${-0.61}^{0.11}_{-0.12}$ & 0.59 $\pm$ 0.17 & ${-0.61}^{0.11}_{-0.12}$ & 0.59 $\pm$ 0.18 \\
NGC\,4631\,c & $-$0.28 $\pm$ 0.01 & 0.88 $\pm$ 0.01 & ${-0.80}^{0.13}_{-0.13}$ & 0.88 $\pm$ 0.05 & ${-0.81}^{0.13}_{-0.13}$ & 0.89 $\pm$ 0.04 \\
NGC\,4631\,d & $-$0.18 $\pm$ 0.06 & 0.96 $\pm$ 0.04 & ${-0.78}^{0.13}_{-0.13}$ & 0.93 $\pm$ 0.05 & ${-0.78}^{0.13}_{-0.13}$ & 0.95 $\pm$ 0.04 \\
NGC\,4631\,e & $-$0.21 $\pm$ 0.03 & 0.94 $\pm$ 0.02 & ${-0.82}^{0.13}_{-0.13}$ & 0.90 $\pm$ 0.04 & ${-0.82}^{0.13}_{-0.13}$ & 0.92 $\pm$ 0.04 \\
NGC\,4631\,Enuc.\,2b & $-$0.12 $\pm$ 0.04 & 0.99 $\pm$ 0.02 & ${-0.78}^{0.13}_{-0.13}$ & 0.97 $\pm$ 0.02 & ${-0.78}^{0.13}_{-0.13}$ & 0.98 $\pm$ 0.02 \\
NGC\,4725\,a & 0.20 $\pm$ 0.34 & 1.00 $\pm$ 0.08 & ${-0.78}^{0.13}_{-0.13}$ & 0.97 $\pm$ 0.02 & ${-0.78}^{0.13}_{-0.13}$ & 0.98 $\pm$ 0.02 \\
NGC\,5194\,Enuc.\,6a & $-$0.10 $\pm$ 0.09 & 1.00 $\pm$ 0.04 & ${-0.78}^{0.13}_{-0.13}$ & 0.95 $\pm$ 0.04 & ${-0.78}^{0.13}_{-0.13}$ & 0.97 $\pm$ 0.03 \\
NGC\,5194\,Enuc.\,3 & $-$0.05 $\pm$ 0.10 & 1.00 $\pm$ 0.04 & ${-0.78}^{0.13}_{-0.13}$ & 0.96 $\pm$ 0.03 & ${-0.78}^{0.13}_{-0.13}$ & 0.97 $\pm$ 0.03 \\
NGC\,5194\,Enuc.\,11a & $-$0.50 $\pm$ 0.18 & 0.66 $\pm$ 0.23 & ${-0.80}^{0.12}_{-0.12}$ & 0.54 $\pm$ 0.27 & ${-0.80}^{0.12}_{-0.13}$ & 0.63 $\pm$ 0.25 \\
NGC\,5194\,Enuc.\,11b & \nodata & \nodata & ${-0.78}^{0.13}_{-0.13}$ & 0.77 $\pm$ 0.18 & ${-0.78}^{0.13}_{-0.13}$ & 0.84 $\pm$ 0.16 \\
NGC\,5194\,Enuc.\,11d & \nodata & \nodata & ${-0.79}^{0.13}_{-0.13}$ & 0.74 $\pm$ 0.23 & ${-0.79}^{0.13}_{-0.13}$ & 0.81 $\pm$ 0.20 \\
NGC\,5194\,Enuc.\,11c & \nodata & \nodata & ${-0.86}^{0.10}_{-0.11}$ & 0.25 $\pm$ 0.21 & ${-0.85}^{0.11}_{-0.11}$ & 0.30 $\pm$ 0.23 \\
NGC\,5194\,c & $-$0.55 $\pm$ 0.09 & 0.59 $\pm$ 0.13 & ${-0.78}^{0.09}_{-0.10}$ & 0.30 $\pm$ 0.21 & ${-0.78}^{0.09}_{-0.10}$ & 0.32 $\pm$ 0.23 \\
NGC\,5194\,Enuc.\,11e & $-$0.26 $\pm$ 0.15 & 0.90 $\pm$ 0.11 & ${-0.77}^{0.13}_{-0.13}$ & 0.82 $\pm$ 0.14 & ${-0.78}^{0.13}_{-0.13}$ & 0.88 $\pm$ 0.11 \\
NGC\,5194\,b & $-$0.62 $\pm$ 0.06 & 0.47 $\pm$ 0.11 & ${-0.77}^{0.07}_{-0.09}$ & 0.22 $\pm$ 0.18 & ${-0.78}^{0.08}_{-0.10}$ & 0.30 $\pm$ 0.21 \\
NGC\,5194\,e & $-$0.76 $\pm$ 0.02 & 0.19 $\pm$ 0.05 & ${-0.80}^{0.02}_{-0.03}$ & 0.05 $\pm$ 0.05 & ${-0.80}^{0.02}_{-0.03}$ & 0.05 $\pm$ 0.05 \\
NGC\,5194\,d & $-$0.84 $\pm$ 0.02 & 0.25 $\pm$ 0.04 & ${-0.89}^{0.02}_{-0.03}$ & 0.04 $\pm$ 0.04 & ${-0.89}^{0.02}_{-0.03}$ & 0.04 $\pm$ 0.04 \\
NGC\,5194\,Enuc.\,1c & $-$0.25 $\pm$ 0.14 & 0.91 $\pm$ 0.10 & ${-0.79}^{0.13}_{-0.13}$ & 0.78 $\pm$ 0.17 & ${-0.79}^{0.13}_{-0.13}$ & 0.84 $\pm$ 0.14 \\
NGC\,5194\,Enuc.\,4a & \nodata & \nodata & ${-0.78}^{0.13}_{-0.13}$ & 0.82 $\pm$ 0.12 & ${-0.78}^{0.13}_{-0.13}$ & 0.92 $\pm$ 0.09 \\
NGC\,5194\,Enuc.\,10a & \nodata & \nodata & ${-0.79}^{0.12}_{-0.12}$ & 0.61 $\pm$ 0.28 & ${-0.79}^{0.13}_{-0.13}$ & 0.79 $\pm$ 0.19 \\
NGC\,5194\,Enuc.\,4c & $-$0.40 $\pm$ 0.12 & 0.77 $\pm$ 0.12 & ${-0.79}^{0.13}_{-0.13}$ & 0.70 $\pm$ 0.18 & ${-0.79}^{0.13}_{-0.13}$ & 0.75 $\pm$ 0.17 \\
NGC\,5194\,a & $-$0.19 $\pm$ 0.09 & 0.95 $\pm$ 0.05 & ${-0.78}^{0.13}_{-0.13}$ & 0.90 $\pm$ 0.07 & ${-0.78}^{0.13}_{-0.13}$ & 0.93 $\pm$ 0.06 \\
NGC\,5194\,Enuc.\,10b & $-$0.30 $\pm$ 0.13 & 0.87 $\pm$ 0.10 & ${-0.77}^{0.13}_{-0.13}$ & 0.74 $\pm$ 0.18 & ${-0.78}^{0.13}_{-0.13}$ & 0.85 $\pm$ 0.12 \\
NGC\,5194\,Enuc.\,4d & $-$0.48 $\pm$ 0.16 & 0.69 $\pm$ 0.20 & ${-0.79}^{0.12}_{-0.12}$ & 0.59 $\pm$ 0.24 & ${-0.79}^{0.12}_{-0.13}$ & 0.67 $\pm$ 0.23 \\
NGC\,5194\,Enuc.\,5a & $-$0.06 $\pm$ 0.15 & 1.00 $\pm$ 0.07 & ${-0.77}^{0.13}_{-0.13}$ & 0.92 $\pm$ 0.06 & ${-0.78}^{0.13}_{-0.13}$ & 0.95 $\pm$ 0.05 \\
NGC\,5194\,Enuc.\,9 & \nodata & \nodata & ${-0.77}^{0.13}_{-0.13}$ & 0.89 $\pm$ 0.10 & ${-0.78}^{0.13}_{-0.13}$ & 0.94 $\pm$ 0.06 \\
NGC\,5194\,Enuc.\,7a & \nodata & \nodata & ${-0.77}^{0.13}_{-0.13}$ & 0.76 $\pm$ 0.21 & ${-0.78}^{0.13}_{-0.13}$ & 0.85 $\pm$ 0.15 \\
NGC\,5194\,Enuc.\,8 & $-$0.33 $\pm$ 0.08 & 0.85 $\pm$ 0.07 & ${-0.74}^{0.13}_{-0.13}$ & 0.80 $\pm$ 0.12 & ${-0.75}^{0.13}_{-0.13}$ & 0.83 $\pm$ 0.10 \\
NGC\,5194\,Enuc.\,7b & $-$0.49 $\pm$ 0.07 & 0.67 $\pm$ 0.09 & ${-0.79}^{0.12}_{-0.13}$ & 0.59 $\pm$ 0.18 & ${-0.81}^{0.12}_{-0.12}$ & 0.66 $\pm$ 0.15 \\
NGC\,5194\,Enuc.\,7c & \nodata & \nodata & ${-0.83}^{0.12}_{-0.12}$ & 0.47 $\pm$ 0.28 & ${-0.82}^{0.12}_{-0.12}$ & 0.54 $\pm$ 0.29 \\
NGC\,5474\, & \nodata & \nodata & ${-0.80}^{0.13}_{-0.13}$ & 0.64 $\pm$ 0.27 & ${-0.79}^{0.12}_{-0.13}$ & 0.74 $\pm$ 0.25 \\
NGC\,5713\,Enuc.\,2a & $-$0.45 $\pm$ 0.03 & 0.73 $\pm$ 0.04 & ${-0.71}^{0.11}_{-0.12}$ & 0.52 $\pm$ 0.18 & ${-0.71}^{0.11}_{-0.13}$ & 0.55 $\pm$ 0.19 \\
NGC\,5713\,Enuc.\,2b & $-$0.23 $\pm$ 0.17 & 0.93 $\pm$ 0.12 & ${-0.85}^{0.10}_{-0.11}$ & 0.25 $\pm$ 0.20 & ${-0.84}^{0.10}_{-0.11}$ & 0.29 $\pm$ 0.22 \\
NGC\,5713\, & $-$0.52 $\pm$ 0.01 & 0.63 $\pm$ 0.02 & ${-0.62}^{0.06}_{-0.13}$ & 0.26 $\pm$ 0.21 & ${-0.60}^{0.05}_{-0.09}$ & 0.24 $\pm$ 0.19 \\
NGC\,5713\,Enuc.\,1 & $-$0.58 $\pm$ 0.01 & 0.54 $\pm$ 0.01 & ${-0.63}^{0.03}_{-0.07}$ & 0.12 $\pm$ 0.12 & ${-0.62}^{0.02}_{-0.04}$ & 0.10 $\pm$ 0.10 \\
NGC\,6946\,Enuc.\,4c & $-$0.23 $\pm$ 0.04 & 0.92 $\pm$ 0.02 & ${-0.72}^{0.14}_{-0.13}$ & 0.90 $\pm$ 0.05 & ${-0.72}^{0.14}_{-0.13}$ & 0.92 $\pm$ 0.05 \\
NGC\,6946\,Enuc.\,8 & $-$0.39 $\pm$ 0.03 & 0.79 $\pm$ 0.03 & ${-0.94}^{0.12}_{-0.12}$ & 0.83 $\pm$ 0.06 & ${-0.95}^{0.12}_{-0.12}$ & 0.84 $\pm$ 0.05 \\
NGC\,6946\,Enuc.\,5a & \nodata & \nodata & ${-0.79}^{0.13}_{-0.13}$ & 0.66 $\pm$ 0.26 & ${-0.79}^{0.13}_{-0.13}$ & 0.81 $\pm$ 0.20 \\
NGC\,6946\,Enuc.\,5b & $-$0.28 $\pm$ 0.20 & 0.89 $\pm$ 0.15 & ${-0.77}^{0.13}_{-0.13}$ & 0.93 $\pm$ 0.05 & ${-0.78}^{0.13}_{-0.13}$ & 0.96 $\pm$ 0.05 \\
NGC\,6946\,Enuc.\,3a & $-$0.50 $\pm$ 0.27 & 0.66 $\pm$ 0.36 & ${-0.78}^{0.13}_{-0.13}$ & 0.93 $\pm$ 0.04 & ${-0.78}^{0.13}_{-0.13}$ & 0.97 $\pm$ 0.03 \\
NGC\,6946\,Enuc.\,3b & $-$0.08 $\pm$ 0.10 & 1.00 $\pm$ 0.05 & ${-0.78}^{0.13}_{-0.13}$ & 0.95 $\pm$ 0.02 & ${-0.78}^{0.13}_{-0.13}$ & 0.98 $\pm$ 0.02 \\
NGC\,6946\,b & $-$0.53 $\pm$ 0.00 & 0.63 $\pm$ 0.01 & ${-0.54}^{0.01}_{-0.02}$ & 0.03 $\pm$ 0.04 & ${-0.53}^{0.01}_{-0.01}$ & 0.03 $\pm$ 0.03 \\
NGC\,6946\,Enuc.\,6b & $-$0.22 $\pm$ 0.11 & 0.93 $\pm$ 0.07 & ${-0.77}^{0.13}_{-0.13}$ & 0.90 $\pm$ 0.08 & ${-0.77}^{0.13}_{-0.13}$ & 0.92 $\pm$ 0.07 \\
NGC\,6946\,Enuc.\,9 & $-$0.07 $\pm$ 0.04 & 1.00 $\pm$ 0.02 & ${-0.77}^{0.13}_{-0.13}$ & 0.98 $\pm$ 0.01 & ${-0.78}^{0.13}_{-0.13}$ & 0.99 $\pm$ 0.01 \\
NGC\,6946\,Enuc.\,7 & $-$0.08 $\pm$ 0.10 & 1.00 $\pm$ 0.05 & ${-0.77}^{0.14}_{-0.13}$ & 0.99 $\pm$ 0.01 & ${-0.78}^{0.13}_{-0.13}$ & 1.00 $\pm$ 0.01 \\
NGC\,6946\,Enuc.\,1 & $-$0.11 $\pm$ 0.08 & 0.99 $\pm$ 0.04 & ${-0.78}^{0.13}_{-0.13}$ & 0.94 $\pm$ 0.05 & ${-0.79}^{0.13}_{-0.13}$ & 0.96 $\pm$ 0.04 \\
NGC\,7331\, & \nodata & \nodata & ${-0.80}^{0.13}_{-0.13}$ & 0.51 $\pm$ 0.29 & ${-0.79}^{0.13}_{-0.13}$ & 0.75 $\pm$ 0.26 \\
\cutinhead{Likely Associated with Supernovae}
NGC\,6946\,Enuc.\,6a & $-$0.19 $\pm$ 0.04 & 0.95 $\pm$ 0.03 & ${-0.81}^{0.13}_{-0.13}$ & 0.91 $\pm$ 0.04 & ${-0.81}^{0.13}_{-0.13}$ & 0.93 $\pm$ 0.04 \\
\cutinhead{Likely AME Candidates}
NGC\,4254\,Enuc.\,1a & \nodata & \nodata & ${-0.78}^{0.13}_{-0.13}$ & 0.88 $\pm$ 0.11 & ${-0.79}^{0.13}_{-0.13}$ & 0.93 $\pm$ 0.07 \\
NGC\,4725\,b & \nodata & \nodata & ${-0.78}^{0.13}_{-0.13}$ & 0.96 $\pm$ 0.03 & ${-0.78}^{0.13}_{-0.13}$ & 0.98 $\pm$ 0.02 \\
NGC\,5194\,Enuc.\,2 & $-$0.23 $\pm$ 0.06 & 0.92 $\pm$ 0.04 & ${-0.79}^{0.13}_{-0.13}$ & 0.89 $\pm$ 0.07 & ${-0.80}^{0.13}_{-0.13}$ & 0.91 $\pm$ 0.06 \\
NGC\,5194\,Enuc.\,1a & 0.05 $\pm$ 0.11 & 1.00 $\pm$ 0.04 & ${-0.78}^{0.14}_{-0.13}$ & 0.95 $\pm$ 0.04 & ${-0.78}^{0.13}_{-0.13}$ & 0.98 $\pm$ 0.03 \\
NGC\,5194\,Enuc.\,1b & 0.02 $\pm$ 0.15 & 1.00 $\pm$ 0.06 & ${-0.78}^{0.13}_{-0.13}$ & 0.95 $\pm$ 0.05 & ${-0.78}^{0.13}_{-0.13}$ & 0.97 $\pm$ 0.03 \\
NGC\,5194\,Enuc.\,4b & $-$0.25 $\pm$ 0.09 & 0.91 $\pm$ 0.07 & ${-0.79}^{0.13}_{-0.13}$ & 0.88 $\pm$ 0.09 & ${-0.79}^{0.13}_{-0.13}$ & 0.90 $\pm$ 0.07 \\
NGC\,6946\,Enuc.\,4a & 0.02 $\pm$ 0.03 & 1.00 $\pm$ 0.01 & ${-0.78}^{0.16}_{-0.14}$ & 0.99 $\pm$ 0.00 & ${-0.78}^{0.13}_{-0.13}$ & 1.00 $\pm$ 0.00 \\
NGC\,6946\,Enuc.\,4b & $-$0.65 $\pm$ 0.11 & 0.42 $\pm$ 0.20 & ${-0.77}^{0.13}_{-0.13}$ & 0.97 $\pm$ 0.01 & ${-0.77}^{0.13}_{-0.13}$ & 0.99 $\pm$ 0.01 \\
NGC\,6946\,a & 0.10 $\pm$ 0.10 & 1.00 $\pm$ 0.03 & ${-0.78}^{0.13}_{-0.13}$ & 0.97 $\pm$ 0.02 & ${-0.78}^{0.13}_{-0.13}$ & 0.99 $\pm$ 0.01 \\
NGC\,6946\,c & $-$0.21 $\pm$ 0.04 & 0.94 $\pm$ 0.02 & ${-0.69}^{0.13}_{-0.13}$ & 0.87 $\pm$ 0.06 & ${-0.73}^{0.13}_{-0.13}$ & 0.91 $\pm$ 0.05 \\
NGC\,6946\,Enuc.\,2a & $-$0.91 $\pm$ 0.43 & 0.24 $\pm$ 0.91 & ${-0.78}^{0.13}_{-0.13}$ & 0.86 $\pm$ 0.06 & ${-0.77}^{0.13}_{-0.13}$ & 0.97 $\pm$ 0.04 \\
NGC\,6946\,Enuc.\,2b & $-$0.10 $\pm$ 0.03 & 1.00 $\pm$ 0.02 & ${-0.78}^{0.13}_{-0.13}$ & 0.98 $\pm$ 0.01 & ${-0.77}^{0.13}_{-0.13}$ & 0.99 $\pm$ 0.01 \\
\enddata
\tablecomments{The uncertainties reported on the thermal fractions for both MCMC fitting scenarios are the scaled MADs. A table with the best-fit amplitudes of the emission components derived from the MCMC fitting scenarios is available in \nameref{app:appendix}.}
\end{deluxetable*}

\subsection{Two-component Fitting}
\label{subsection:lsq}

Fitting a power law to the measured flux densities ($S_{\nu} \propto \nu^{\alpha}$, where $S_{\nu}$ is the flux density at a given frequency $\nu$, and $\alpha$ is the corresponding index of the power law) is the simplest way to model the radio spectra. Splitting a power law fit into two components allows us to decompose the spectra into its synchrotron (3 to 33\,GHz) and free-free (33 to 90\,GHz) contributions \citep{Condon1992}. This model sufficiently describes the primary emission mechanisms at radio frequencies for a majority of the star-forming galaxies in the local universe. Namely, by including 90\,GHz data in the fitting, we anticipate learning more about the impact of thermal dust, which is expected to contribute at such high (i.e., $\gtrsim$ 30\,GHz) frequencies. A power law model was applied to our observations using the Python package \texttt{scipy.curve\_fit}, which performs a non-linear least squares fit to the flux densities. Fitting was performed only for sources with significant detections (S/N $>$ 3) in at least two bands, for both the 3 to 33\,GHz and 33 to 90\,GHz regimes. 

\begin{figure}[htp]
\includegraphics[width=0.45\textwidth]{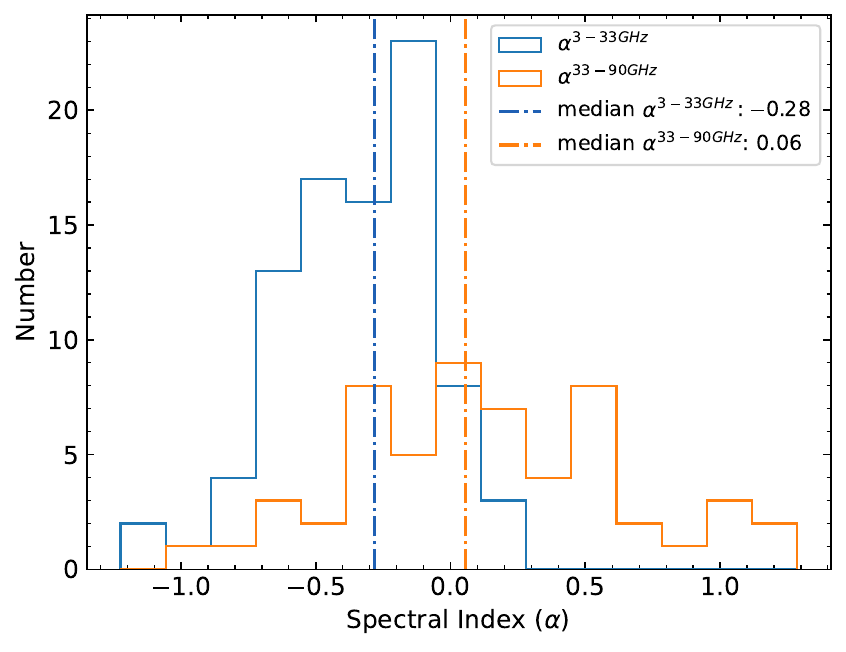}
    \caption{Spectral index distributions for two-component power law fitting for 3 to 33 and 33 to 90\,GHz. There are clear differences in the peaks and distribution widths; the distribution of the spectral indices measured from 3 to 33\,GHz is narrower and peaks at $-0.028$, while the distribution of the spectral indices measured from 33 to 90\,GHz is more extended and peaks at 0.06. This demonstrates the need for at least one additional power law component in order to adequately characterize the observed radio spectra.}
    \label{fig:spectralindexpowerlaw}
\end{figure}

First, a single power law was fit to the spectra to calculate spectral indices for 90 regions (19 nuclear and 71 extranuclear) between 3 to 90\,GHz, resulting in a median $\alpha$ of $-0.21 \pm 0.03$ with a scatter of 0.24. The data were then split into low and high frequency regimes, and the power law was modified to fit two components. Spectral indices were calculated for 87 regions (19 nuclear and 68 extranuclear regions) for 3 to 33\,GHz, resulting in a median $\alpha$ of $-0.28 \pm 0.04$ with a scatter of 0.29. Similarly, for 56 regions (15 nuclear and 41 extranuclear) for 33 to 90\,GHz, we find a median $\alpha$ of $0.06 \pm 0.09$ with a scatter of 0.55. Figure \ref{fig:spectralindexpowerlaw} shows the distributions of spectral indices for 3 to 33 and 33 to 90\,GHz, with the median values for both regimes. These results indicate that the high-frequency emission component of the observed spectra is flatter, suggesting that there is spectral curvature and necessitating a more complex (i.e., at least two component) model to properly describe the radio spectra. 

Following \citet{Klein1984}, thermal fractions at 33\,GHz are calculated such that,

\begin{equation}
    f_{\mathrm{T}}^{\nu_{1}}=\frac{ (\frac{\nu_{2}}{\nu_{1}})^{-\alpha} - (\frac{\nu_{2}}{\nu_{1}})^{-\alpha^{\mathrm{NT}}}} { (\frac{\nu_{2}}{\nu_{1}})^{-0.1} - (\frac{\nu_{2}}{\nu_{1}})^{-\alpha^{\mathrm{NT}}}}
    \label{eq:tfrac}
\end{equation} where $\nu_{1}$ is the target frequency (33\,GHz), $\alpha$ is the observed index from 3 to 33\,GHz, and $\alpha^{\mathrm{NT}}$ is the non-thermal (synchrotron emission) spectral index. Previous literature reports an expected value for $\alpha^{\mathrm{NT}}$ of $-0.83$ and a scatter of 0.13 \citep{Niklas1997,Murphy2011,Linden2020}. Based on these results, for calculating thermal fractions from the power law fitting, the value of $\alpha^{\mathrm{NT}}$ was fixed to $-0.83$ and $\alpha$ was taken as the spectral index measured from 3 to 33\,GHz for all cases where $\alpha$ was measured to be $>$ $-0.83$. For cases where the measured $\alpha$ was $<$ $-0.83$, we subtracted a single standard deviation (i.e., 0.13) from $\alpha$ and set $\alpha^{\mathrm{NT}}$ as the resulting value. We find a median thermal fraction at 33\,GHz of 88 $\pm$ 2\% with a scatter of 17\%.

\subsection{MCMC Fitting}
\label{mcmc}

While the basic power law model informs us that at least two emission components are making up the observed radio spectra, thermal dust potentially also affects the data at 33\,GHz, presenting an additional component to which our two-component power law model is insensitive. A total of 46 of the 119 regions in our sample show 90\,GHz emission that is at least twice as great as the 33\,GHz emission, indicating the potential presence of thermal dust emission.  Including the newly acquired 90\,GHz data allows us to use a more sophisticated model to determine how much of the emission at 33\,GHz may be impacted by the presence of thermal dust.    

For 119 sources (23 nuclear and 96 extranuclear regions), we performed MCMC fitting based on the equations that describe the physical processes producing the observed spectra. We chose MCMC fitting based on its advantages of more accurate estimates of uncertainties on parameters and its ability to thoroughly sample a multidimensional parameter space. We perform fitting using the Python package \texttt{emcee} \citep{emcee} to produce posterior probability distributions for four fitted parameters: the non-thermal spectral index ($\alpha^{\mathrm{NT}}$) and the amplitudes of the synchrotron, free-free, and thermal dust emission ($A_{s}$, $A_{ff}$, and $A_{d}$, respectively). Note that we did not fit for AME despite its potential presence (see \citet{Hensley2015} and \citet{Murphy2020} for examples of fitting performed on extragalactic radio spectra that do take AME into account). Fitting for AME is beyond the scope of this work and will be covered in future work. 

The thermal dust emission is modeled as a modified blackbody, i.e., 

\begin{equation}
    S_\nu^{d} = A_d \left( \frac{\nu}{353\,{\rm GHz}} \right)^\beta B_\nu\left(T_d\right)
\end{equation}

\noindent
where $A_{d}$ is the amplitude of the dust emission at the reference frequency of 353\,GHz, $T_{d}$ is the dust temperature, and $B_{\nu}(T)$ is the Planck function. We fix $T_d$ = 20 K, and $\beta$ = 1.5 in accordance with the procedure followed in \citet{Murphy2020}. 

The free-free emission is modeled as

\begin{equation}
    S_\nu^{\mathrm{ff}} = A_{ff} \frac{g_{ff}\left(\nu, T\right)}{g_{ff}\left(30\,{\rm GHz}, T\right)}
\end{equation}

\noindent
where $A_{ff}$ is the amplitude of the free-free emission and $g_{ff}$ is the Gaunt factor. The Gaunt factor is approximated as a function of frequency and electron temperature, i.e., 

\begin{equation} 
    g_{\mathrm{ff}}(\nu,T_{e})=\mathrm{ln  \biggl\{ exp  \Bigl[ 5.960 - \frac{\sqrt{3}} {\pi} ln} (\nu_{9}T_{4}^{-3/2}) \Bigr] + e \biggr\}
\end{equation} where $\nu_{9} \equiv \nu/10^{9}$ Hz and $T_{4} \equiv T_{e}/10^{4}$ K. $T_{e}$ is fixed to $10^{4}$ K in accordance with the procedure followed in \citet{Murphy2020}. 

The synchrotron emission is modeled as

\begin{equation}
    S_\nu^{\mathrm{sync}} = A_s \left( \frac{\nu}{30\,{\rm GHz}} \right)^{\alpha^{\mathrm{NT}}}
\end{equation} where $A_{s}$ is the 30 GHz flux density in the optically thin limit and $\alpha^{\mathrm{NT}}$ is the non-thermal spectral index. 
We require that all amplitudes of the emission components (i.e. $A_s,A_{ff}$ and $A_d$) are positive. The derived best-fit amplitudes per source for both MCMC fitting scenarios are given in \nameref{app:appendix}. We also implemented a Gaussian prior on $\alpha^{\mathrm{NT}}$ with an expected value of $-0.83$ and a scatter of 0.13, based on results from previous investigations \citep{Niklas1997,Murphy2011,Linden2020}. Without this Gaussian prior, we find a median non-thermal spectral index of $-0.33 \pm 0.06$ with a scatter of $0.55$ for fitting with a dust component and 90\,GHz data, and $1.95 \pm 0.45$ with a scatter of $3.90$ for fitting without a dust component and with only VLA data. It is clear that without this prior, the resulting non-thermal spectral indices can become unphysical (e.g., highly positive), which is inconsistent with expectations for known acceleration mechanisms for cosmic-ray electrons in the interstellar medium \citep{Longair1994}.

\begin{figure*}[htp]
    \centering
\includegraphics[width=1.0\textwidth]{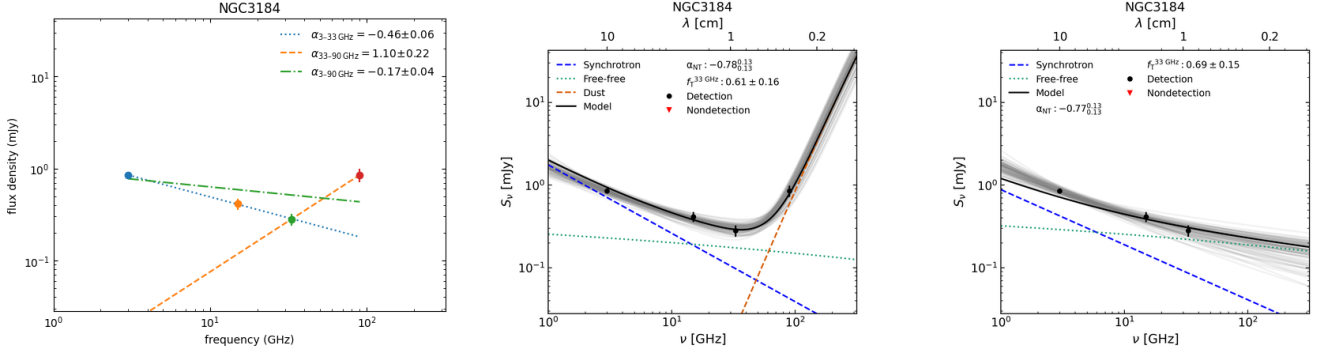}
    \caption{Example of MCMC fitting performed for NGC\,3184. Left: fitting with power laws. Middle: fitting with VLA and GBT data and four parameters. Right: fitting with only VLA data and three parameters, excluding fitting for the thermal dust component. The blue dashed lines, green dotted lines, and orange dashed line represent the contributions from synchrotron, free-free, and thermal dust emission. The gray lines represent 100 randomly selected models produced by the MCMC fitting as a visualization of the uncertainty on our best-fit model. This is an instance of a source with a flux density at 90 GHz that is 2$\sigma$ greater than the flux density at 33 GHz, making it a strong candidate for showing a potential impact of including thermal dust and 90 GHz in MCMC fitting. We investigate all of these cases in our sample in further detail in Section \ref{subsubsection:comparison}. For this source, the value of $\alpha^{\mathrm{NT}}$ does not change significantly between the two fitting scenarios, but we do find a greater thermal (free-free) fraction for fitting without dust and GBT data, matching the overall trend we observe. A corresponding version of this figure for all other sources is available in the online journal as a figure set (119 images).}
 
    \label{fig:mcmcfitting}
\end{figure*}

\begin{figure}[htp]
\includegraphics[width=0.45\textwidth]{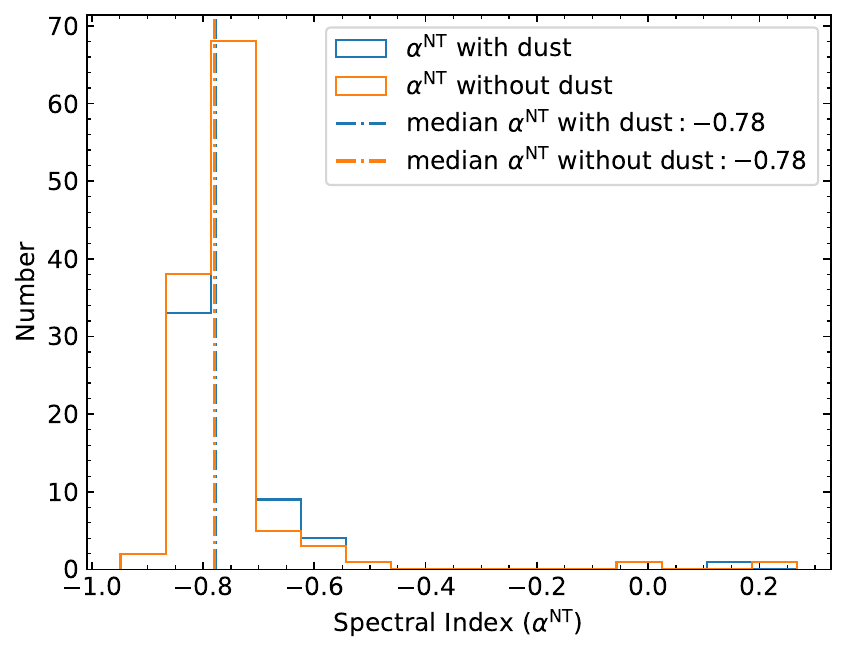}
    \caption{Spectral index distributions for MCMC fitting performed with (blue line) and without (orange line) accounting for dust. We observe a tight distribution with a small scatter especially when compared to Figure \ref{fig:spectralindexpowerlaw}. The distributions of $\alpha^{\mathrm{NT}}$ for both fitting scenarios are basically identical; the median difference between the two distributions is 0.001. This is almost certainly due to the prior imposed on the value of $\alpha^{\mathrm{NT}}$, which is required in order to get physically feasible results from the MCMC fitting.}
\label{fig:spectralindexmcmc}
\end{figure}

The final model spectrum is the sum of these three terms. We make 300,000 realizations of this model twice per source, first without precise initial guesses and then with the results from the first iteration as inputs for the initial guesses. This repetition increased the accuracy of the final estimated parameters. This process was used for fitting both with VLA and GBT data and taking thermal dust emission into account, as well as with only VLA data and without taking dust emission into account. Figure \ref{fig:mcmcfitting} shows an example of MCMC fitting both with and without dust for a representative sample source (NGC\,3184). Corresponding versions of this figure, with the power law fitting results included, are shown for all sources in the online journal as a figure set (see Figure \ref{fig:mcmcfitting}).
We find a median non-thermal spectral index of $-0.78 \pm 0.002$ with a scatter of 0.02 for fitting VLA and GBT data with dust, while we find a median non-thermal spectral index of $-0.78 \pm 0.002$ with a scatter of 0.01 for fitting only VLA data without dust. The narrow distributions and small median difference between the two fitting methods are almost certainly due to the prior imposed on the value of the non-thermal spectral index. As stated in Section \ref{mcmc}, setting the chosen prior is necessary in order to produce reasonable, physical values for the non-thermal spectral indices; finer spectral coverage, especially at lower frequencies, could allow for this prior to be relaxed. Figure \ref{fig:spectralindexmcmc} shows the distributions of spectral indices for MCMC fitting both with and without dust, with the median values for both scenarios.

\begin{figure}[htp]
    \includegraphics[width=0.45\textwidth]{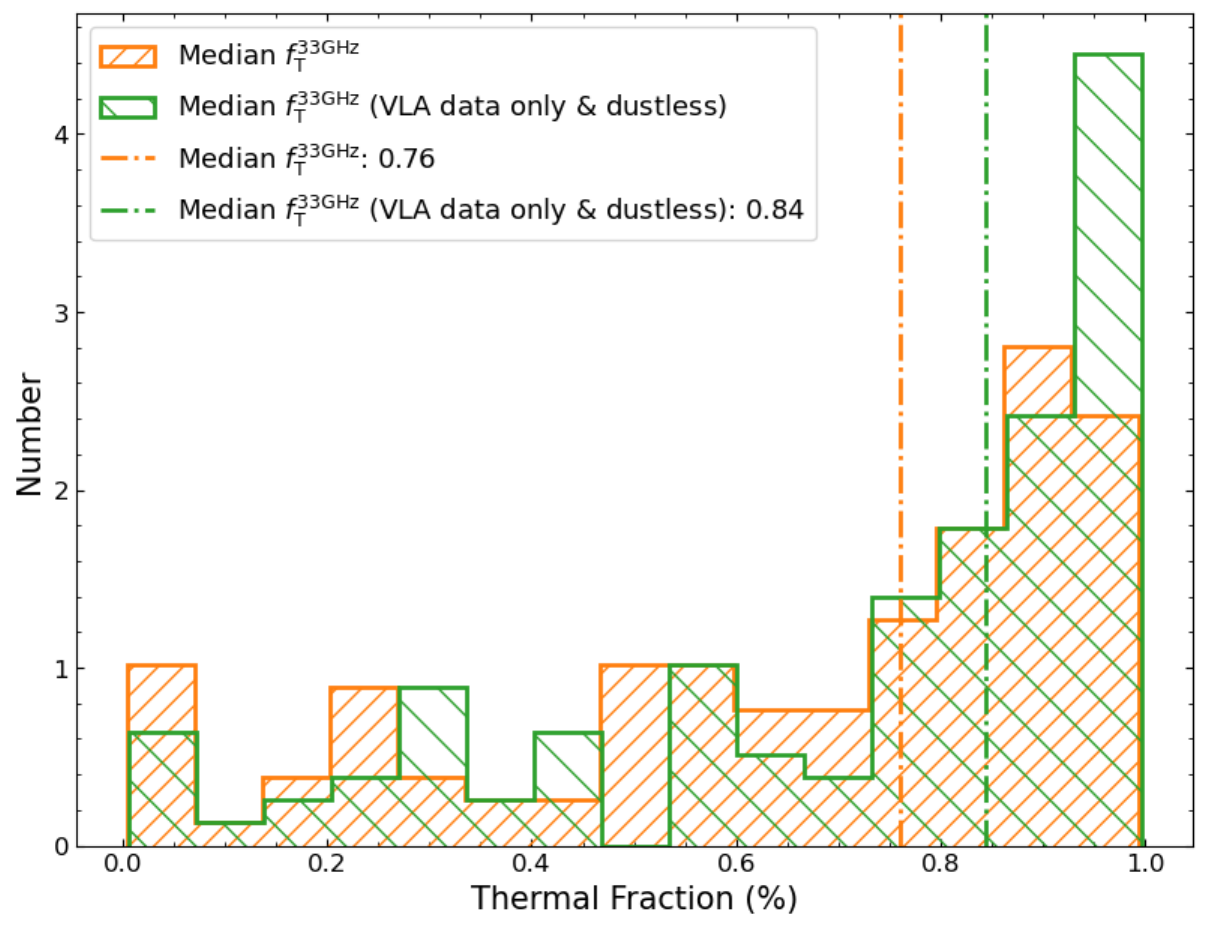}
    \caption{Histograms of the thermal (free-free) fractions calculated for MCMC fitting performed on both VLA and GBT data and with fitting for thermal dust (orange line), as well as for MCMC fitting performed on only VLA data and without fitting for thermal dust (green line). The distributions of the thermal fractions for both fitting scenarios overlap significantly; the median difference between the two distributions is $4.91\%$. However, the median thermal fraction for fitting without dust and GBT data is slightly higher than fitting for thermal dust and with GBT and VLA data.}
    \label{fig:thermalfracs}
\end{figure}

Similarly, we calculated thermal fractions based on the results from the two different fitting scenarios by dividing the free-free model by the total emission model for each of the 300,000 realizations and taking the median of these 300,000 ratios to be the best-fit final thermal fraction. We report the scaled MAD of all 300,000 models as the uncertainty on the best-fit thermal fraction. We find the median thermal fraction at 33\,GHz to be $76 \pm 3$\% with a scatter of 25\% for fitting VLA and GBT data with dust, while we find the median thermal fraction at 33\,GHz to be $84 \pm 2$\% with a scatter of 18\% for fitting only VLA data without dust. We find a median ratio of thermal fractions for fitting with and without dust of ${<\frac{f_{\mathrm{T}}}{f_{\mathrm{T,dustless\ VLA}}}>} \approx 0.91$ with a scatter of 0.06. The median difference in thermal fractions per source between both MCMC fitting methods is $4.91\%$.
Figure \ref{fig:thermalfracs} shows the distributions of thermal fractions for MCMC fitting both with and without dust, with the median values for both scenarios. 

\noprint{\figsetstart}
\noprint{\figsetnum{3}}
\noprint{\figsettitle{Fits}}

\figsetgrpstart
\figsetgrpnum{3.1}
\figsetgrptitle{NGC3184}
\figsetplot{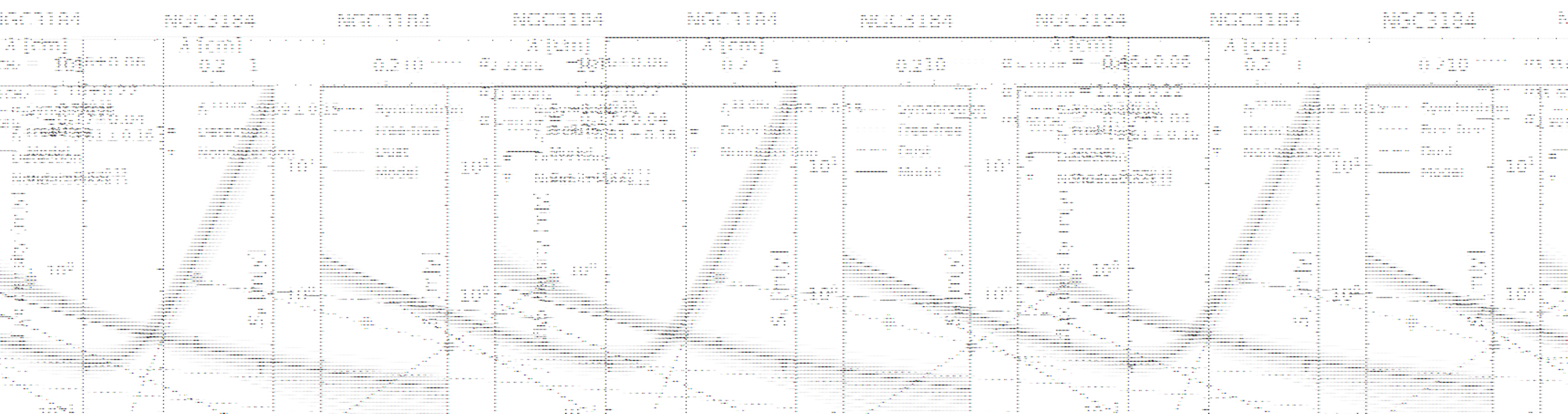}
\figsetgrpnote{Power law and MCMC fits for the corresponding region}
\figsetgrpend

\figsetend

\subsection{Impact of Local Background Subtraction}

To investigate the sensitivity of our results to the presence of local backgrounds, we repeated the aperture photometry after subtracting local background measurements on: (i) all four bands and (ii) only 90\,GHz flux densities. For the 3 to 33\,GHz power law fitting, we find thermal fractions of (i) 87 $\pm$ 3\% with a scatter of 19\% and (ii) 86 $\pm$ 3\% with a scatter of 20\%. For the MCMC fitting with thermal dust and 90\,GHz data, we find thermal fractions of (i) 74 $\pm$ 3\% with a scatter of 26\% and (ii) 85 $\pm$ 2\% with a scatter of 17\%. 
Because these values are similar to what is measured without removing a local background, as well as comparable to one another given their respective scatters, we conclude that local background subtraction does not affect our results.  

\section{DISCUSSION}
\label{section:DISCUSSION}

In the following section we discuss the implications of our results, including their implications for measuring star formation rates using radio continuum data.

\subsection{Comparison of Fitting Methods}
\label{subsection:comparison}

Below we discuss the results from our different fitting methods.

\subsubsection{Power Law Fitting}

The difference in the distributions of spectral indices for low and high frequency regimes found using two-component power law fitting is evidence that a single power law fit is not sufficient enough to describe the observed radio spectra. While the distribution of spectral indices measured from 3 to 33\,GHz peaks at ${-}0.28 \pm 0.04$ with a scatter of 0.29, the distribution of spectral indices measured from 33 to 90\,GHz is significantly flatter, peaking at $0.06 \pm 0.09$ with a scatter of 0.55. The distribution of spectral indices measured for 3 to 90\,GHz peaks in between these two distributions at $-0.21 \pm 0.03$ with a scatter of 0.24. By splitting a traditional power law fit into two components, the distinct low and high frequency regimes of the observed spectra are more accurately characterized, corresponding to the different physical emission processes which dominate below and above $\sim$\,33\,GHz.

\subsubsection{MCMC Fitting}

We performed MCMC fitting for two scenarios: fitting VLA and GBT data and accounting for thermal dust emission versus fitting only VLA data and without accounting for thermal dust emission. This allows us to study the significance of including 90\,GHz data and thermal dust in the decomposition of radio spectra (i.e., to determine how much dust contributes to the observed 33\,GHz emission) and, ultimately, the estimation of the amount of free-free emission for calculating star formation rates. We find that the inclusion of 90\,GHz and thermal dust in MCMC fitting produces a median thermal fraction at 33\,GHz of 76 $\pm$ 3\% with a scatter of 25\%, while the exclusion of 90\,GHz and thermal dust produces a median thermal fraction at 33\,GHz of 84 $\pm$ 2\% with a scatter of 18\%.

While the inclusion of the GBT data and thermal dust component does result in a smaller estimated thermal fraction at 33\,GHz, the difference remains within the 1$\sigma$ scatter of the measurements indicating that dust does not contribute significantly to the 33\,GHz emission. This matches with previous observations of the Milky Way that show that free-free is the dominant emission mechanism at 33\,GHz \citep{Planck2015}. The bottom panels of Figure \ref{fig:trends} show the distribution of thermal fraction ratios, or the ratios of the thermal fractions calculated from the two MCMC fitting methods, for each region in the sample as a function of galactocentric radius ($r_{\mathrm{G}}$) and photometric aperture diameter ($d_{\mathrm{ap}}$). We also investigated the thermal fraction ratio as a function of 3 and 33\,GHz flux densities and found no obvious trend between either relationship. It is clear that the vast majority of the measured thermal fractions do not vary significantly ($>1\sigma$) with the MCMC fitting method, i.e., the potential presence of thermal dust emission does not dominate the radio emission at 33\,GHz.

The MCMC fitting results (i.e., non-thermal spectral index and thermal fraction values) for both scenarios for NGC\,4579 and NGC\,4594\,a differ significantly from the respective median non-thermal spectral indices. This is most likely due to the presence of an AGN in the nucleus of both galaxies influencing the measured 33\,GHz flux densities, as discussed by \citet{Murphy2018}, who found NGC\,4579 and NGC\,4594 to be outliers when comparing the ratio of 33\,GHz flux to H$\alpha$ line flux versus galactocentric radius. 

\subsubsection{Power Law versus MCMC Fitting} \label{subsubsection:comparison}

The inclusion of thermal dust and 90\,GHz data is found to marginally affect the calculated thermal fractions, as evidenced in the difference between the median values calculated for the two fitting methods. While we find a median thermal fraction of $88 \pm 2$\% with a scatter of 17\% with 3 to 33\,GHz power law fitting, we find a median thermal fraction of $76 \pm 3$\% with a scatter of 25\% for MCMC fitting when taking both VLA and GBT data and thermal dust into account. However, we find a median thermal fraction of $84 \pm 2$\% with a scatter of 18\% when performing MCMC fitting on only VLA data and without thermal dust. We conclude that the thermal fractions at 33\,GHz found using MCMC fitting without dust and 90\,GHz data are not a significant improvement on the thermal fractions calculated from simply the 3 to 33\,GHz power law fitting. However, MCMC fitting is advantageous over power law fitting in that it produces more reliable error distributions because it is sampling a larger parameter space. The median values and scaled MADs of the thermal fractions and spectral indices from these three fitting methods are summarized in Table \ref{tab:summary}.

\begin{deluxetable*}{l|cccc}
\tablecaption{Summary of Derived Parameters \label{tab:summary}} 
\tablehead{
\colhead{Fitting method} &
\colhead{med($\alpha$)} &
\colhead{MADN($\alpha$)} &
\colhead{$f_{\mathrm{T}}^{\mathrm{33\,GHz}}$} &
\colhead{MADN($f_{\mathrm{T}}^{\mathrm{33\,GHz}}$)}
}
\startdata
3\textendash33\,GHz power law & $-0.28 \pm 0.040$ & 0.29 & $0.88 \pm 0.023$ & 0.17 \\
MCMC w/ dust \& 90\,GHz & $-0.78 \pm 0.002$ & 0.02 & $0.76 \pm 0.028$ & 0.25 \\
MCMC w/o dust \& 90\,GHz & $-0.78 \pm 0.002$ & 0.01 & $0.84 \pm 0.021$ & 0.18 \\
\enddata
\tablecomments{$\alpha =  \rm{\alpha^{NT}} $ for the 3 to 33\,GHz power law fitting method; $\alpha = \rm{\alpha^{3\textendash33\,GHz}}$ for both MCMC fitting methods. MADN$=$scaled MAD.}
\end{deluxetable*}

We also investigated the impact of regions where the 90\,GHz flux density is at least 2$\sigma$ greater than the 33\,GHz flux density, indicating the potential presence of thermal dust. Focusing on these 46 cases out of 119 total regions, we find a median thermal fraction through 3 to 33\,GHz power law fitting of $73 \pm 7$\% with a scatter of 33\%, a median thermal fraction through MCMC fitting with thermal dust and 90\,GHz data of $56 \pm 9$\% with a scatter of 50\%, and a median thermal fraction through MCMC fitting without thermal dust and 90\,GHz data of $75 \pm 6$\% with a scatter of 33\%. Conversely, by excluding these 46 sources, we find median thermal fractions through the same three respective fitting methods of $90 \pm 3$\% with a scatter of 15\%, $80 \pm 3$\% with a scatter of 17\%, and $86 \pm 2$\% with a scatter of 13\%. We find the greatest differences in median thermal fractions for MCMC fitting with thermal dust and 90\,GHz data included; by excluding our subset of 46 sources, the calculated median thermal fraction increases by 4\%. We conclude that sources with larger 90\,GHz flux densities result in lower thermal fractions for all three fitting methods, but the respective scatters for this subset are so large that the thermal fractions are still within 1$\sigma$ of each other, indicating that these regions harboring dust do not significantly impact our results. 

\subsection{Implications for Star Formation Rate Estimates}

As previously mentioned, the median ratio of the thermal fractions of all 119 regions for fitting with and without dust is ${<\frac{f_{\mathrm{T}}}{f_{\mathrm{T,dustless\ VLA}}}>} \approx 0.91$ with a scatter of 0.06. In other words, the thermal fractions derived from fitting without dust and 90\,GHz are on average 10\% larger than those found from fitting with dust and 90\,GHz data. 
When the supernova-associated source and all AGN and SF/AGN sources with $r_{\mathrm{G}}<500$\,pc are removed according to the designations in Table \ref{tab:properties}, so that we are left with only 103 pure star-forming regions, we find a median ratio of thermal fractions between the two MCMC fitting methods of ${<\frac{f_{\mathrm{T}}}{f_{\mathrm{T,dustless\ VLA}}}>} \approx 0.94$ with a scatter of 0.06. This suggests that star formation rates based on estimating 33\,GHz thermal fractions using lower frequency radio data alone are robust, and would only require a $\approx5\%$ correction, on average,to account for the decrease in thermal fraction when higher frequency data are included in the spectral decomposition.

\subsection{Comparison to Previous Work}

While \citet{Linden2020} found a slightly higher median thermal fraction at 33\,GHz for 163 regions at 7\arcsec \,resolution of $94 \pm 1$\% with a smaller scatter of 8\%, they performed aperture photometry with an aperture diameter of 7\arcsec, corresponding to a median aperture diameter of $259 \pm 7$\,pc.  
Photometry in the present analysis was performed on 10\arcsec \,resolution images with a median aperture diameter of $800 \pm  14$\,pc. 
Thus, our measured thermal fraction of $88 \pm 2$\% with a scatter of 17\% follows the expectation that our larger aperture sizes contain more diffuse (predominantly non-thermal) emission, resulting in smaller thermal fractions. This observed decrease in thermal fraction is consistent with the expectation that by convolving from 7 to 10\arcsec, the highly-peaked free-free emission is more distributed (``smeared out'') leading to smaller free-free fractions.
Consequently, our results demonstrate that on $\sim800\,$pc scales the 33\,GHz data is still largely dominated by free-free emission even when additionally accounting for a contribution from thermal dust.  
This suggests that 33\,GHz flux densities should provide robust estimates of massive star formation rates for H\,II and star-forming regions on $800\,$pc scales within galaxies.

Our results also align with those of \citet{Emig2020}; for 27 candidate super star clusters in NGC\,4945 at 2.2\,pc resolution, they found that dust is faint enough at this frequency that it does not have a large impact on their calculated thermal fractions at 93\,GHz, the median of which was 62\%. Moreover, \citet{Mills2021} studied the central 200\,pc of NGC\,0253 at 94\,GHz, calculating thermal fractions at 3\,mm ranging from 79\% to 99\% for four isolated sources detected in the central region at a resolution of 3.8\,pc.  

\subsubsection{Galactocentric Radius and Aperture Diameter}
\label{subsubsection:trends}

\begin{figure*}[htp]
\includegraphics[width=1.0\textwidth]{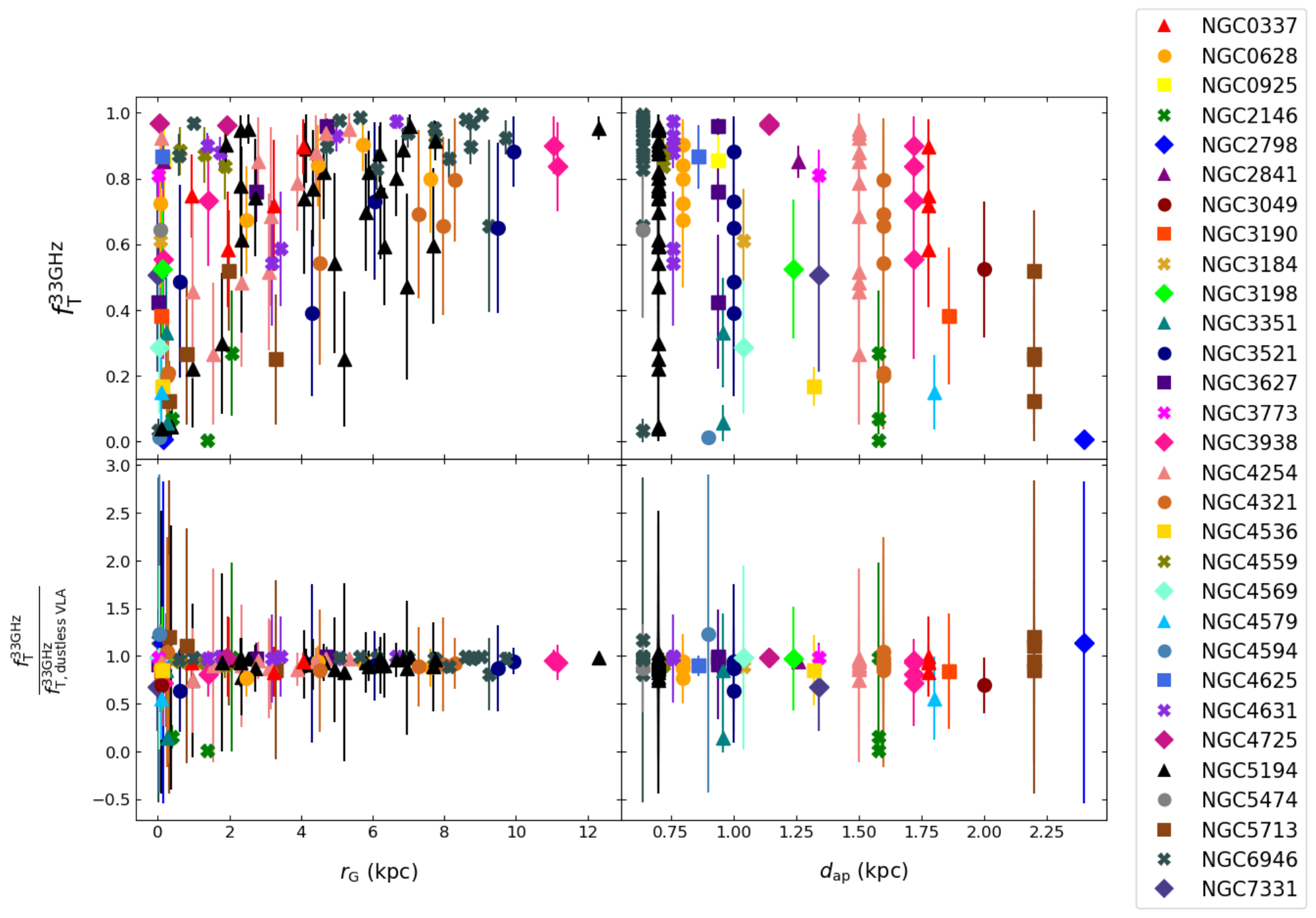}
    \caption{Top panels: distributions of the thermal (free-free) fractions found using the four-parameter model that included 90\,GHz data and thermal dust versus galactocentric radius and photometric aperture diameter. Bottom panels: distributions of the thermal fraction ratios for the four-parameter model versus the three-parameter model that excluded 90\,GHz data and thermal dust versus galactocentric radius and photometric aperture diameter. We note low thermal fractions for two of the three regions associated with NGC\,2146. Sample size: 118 regions (excluding the source likely associated with a supernova, NGC\,6946\,Enuc.\,6a).}
    \label{fig:trends}
\end{figure*}

Following the procedure in \citet{Linden2020}, we apply a radial cut ($r_\mathrm{G}$\,$\sim$\,250 pc) to the results from MCMC fitting with VLA and GBT data and thermal dust and remove the source likely associated with a supernova, splitting the sample into 23 nuclear regions and 95 extranuclear regions. We find the median thermal fraction at 33\,GHz to be $49 \pm 11$\% with a scatter of 43\% for nuclear regions, while we find a median thermal fraction at 33\,GHz of $80 \pm 3\%$ with a scatter of 21\% for extranuclear regions. This agrees with the finding in \citet{Linden2020} that the median thermal fraction at 33\,GHz decreases and the scatter increases after limiting the sample to nuclear regions with $r_\mathrm{G}$ $<$ 250 pc. Qualitatively, this also matches the observed decrease in spread of thermal fractions as galactocentric radius increases (top left panel of Figure \ref{fig:trends}). This indicates that there is a greater non-thermal emission component for the circumnuclear star-forming regions studied here.

Similarly, we apply a cut based on the median aperture diameter ($d_{\mathrm{ap}}=800\,\mathrm{pc}$) to the MCMC results and split the sample into 60 ($d_{\mathrm{ap}}<800\,\mathrm{pc}$) and 58 ($d_{\mathrm{ap}}\geq800\,\mathrm{pc}$) regions. For the 60 regions, we find the median thermal fraction at 33\,GHz to be 76 $\pm$ 2\% with a scatter of 15\%. We find the median thermal fraction at 33\,GHz to be 56\% $\pm$ 7\% with a scatter of 41\% for the 63 regions. From this we conclude that on $\gtrsim$0.8\,kpc scales, an increase in the diameters of the photometric apertures produces a decrease in the measured thermal fractions (see the top right panel of Figure \ref{fig:trends} for the distribution of thermal fractions against aperture diameter). 

This result also matches the results from \citet{Linden2020}, who found thermal fractions as low as 40 to 50\% for photometric apertures larger than 1 kpc, whereas apertures smaller than 1 kpc produced thermal fractions as high as 90\%. This is as expected, since larger apertures contain more non-thermal emission relative to thermal free-free emission, which is typically compact and centered on the H\,II region.   

One galaxy, NGC\,2146, in Figure \ref{fig:trends} (green markers) stands out compared to all the others. We find lower thermal fraction ratios versus $r_\mathrm{G}$ and $d_\mathrm{ap}$ for two of the three regions associated with NGC\,2146 (the nuclear region b and the extranuclear region c). NGC\,2146 is a dusty, highly star-forming Luminous Infrared Galaxy \citep{Kennicutt2011} and is the most IR-luminous galaxy in the sample. As mentioned in Section \ref{section:PHOTOMETRY AND ANALYSIS}, 15\,GHz flux densities for these regions were excluded from the fitting, however, it is unlikely that this is a primary contributing factor for this difference.   

\section{CONCLUSIONS}
\label{section:CONCLUSIONS}

We have presented new 90\,GHz continuum imaging of 119 star-forming regions (23 nuclear and 96 extranuclear) in 30 nearby SFRS galaxies and combined it with previous 3, 15, and 33\,GHz VLA imaging.  Of these, 107 are identified as star-forming regions, 1 is a likely supernova remnant, and 12 are likely AME candidates. We have decomposed the radio spectra of these regions into synchrotron, free-free, and thermal dust emission components through power law fitting from 3 to 33\,GHz, as well as through MCMC fitting for two scenarios: with thermal dust and 90\,GHz data, and without thermal dust and 90\,GHz data. For these three fitting instances, we computed the non-thermal spectral indices and thermal (free-free) fractions at 33\,GHz.

From this analysis, we find the following:

\begin{enumerate}
    \item We first fit a simple power law to all of the measured flux densities, i.e., between 3 and 90\,GHz. We then fit two power laws to the data between 3 and 33 and 33 to 90\,GHz. From this, we find that at least two components are needed to adequately describe our radio spectra into non-thermal and thermal emission. We find a median spectral index from 3 to 33\,GHz of $-0.28 \pm 0.04$ with a scatter of 0.29. Using this result, we calculate the median thermal (free-free) fraction at 33\,GHz to be $88 \pm 2$\% with a scatter of 17\%.
    
    \item We evaluated two scenarios of MCMC fitting: (i) with all of the data and including a thermal dust emission component and (ii) excluding both the 90\,GHz data and a thermal dust component. The first scenario produces a median thermal (free-free) fraction of $76 \pm 3$\% with a scatter of 25\%, while the second scenario produces a median thermal fraction of $84 \pm 2$\% with a scatter of 18\%. From this, we conclude that free-free emission, not thermal dust, dominates the observed 33\,GHz emission on the physical scales studied here (i.e., $\sim$800\,pc). 

   \item While thermal (free-free) emission fractions are $\approx$10\% larger when excluding thermal dust in the MCMC fitting, removing sources that are likely contaminated by AGN decreases this discrepancy to only 5\%. Consequently, star formation rates measured by estimating thermal fractions at 33\,GHz are only mildly biased high in the absence of having 90\,GHz data when decomposing radio spectra.
    
    \item We do not find a noticeable difference between a simple power law fit from 3 to 33\,GHz versus MCMC fitting for calculating thermal (free-free) fractions of the same data set; that is, MCMC fitting does not perform better than power law fitting for the radio spectra in our sample. This is consistent with the findings of \citet{Linden2020}.
    
    \item The median thermal (free-free) fraction measured through power law fitting from 3 to 33\,GHz matches that found by \citet{Linden2020}, falling within 1$\sigma$ of each other, where $\sigma$ is the respective scatters of 8\% and 17\% added in quadrature. This overlap is expected when differences in aperture sizes are taken into account. We also validate their finding that the median thermal fraction decreases while the scatter in thermal fractions increases towards the centers of the galaxies in our sample. Specifically, we find a decrease in the median thermal fraction by a factor of $\sim$\,1.3 and an increase in scatter by a factor of $\sim$\,1.8 for nuclear regions.
\end{enumerate}

Looking forward, the next generation Very Large Array \citep[ngVLA;][]{ngvla} and Square Kilometre Array \citep[SKA;][]{ska} will have the advantages of greater sensitivity and wider bandwidths, revealing fainter H\,II regions and allowing us to more accurately measure spectral indices in single bands rather than requiring multiple observations. Both facilities will also increase the number of observable galaxies in their free-free emission. With the ngVLA, we will acquire even higher frequency data (90\,GHz), whereas data from the SKA will cover the sub-GHz regime to definitively determine $\alpha^{\mathrm{NT}}$ and continue filling in the mid- to high-frequency range ($\gtrsim$ 50\,GHz). Overall, the increased sensitivity and wider bandwidths of the ngVLA and SKA will greatly increase the sample size of H\,II regions and further constrain the amount of pure free-free emission present in them. By extending our access to much fainter regions in nearby galaxies and global measurements of galaxies at high redshift, these facilities will help  improve the current census of massive star formation rates. Our results have demonstrated that 33\,GHz emission can act as a robust estimate for star formation rates in these galaxies that will be observed in future studies with these facilities.

\begin{acknowledgments}
A.R.D. was supported by the NRAO/GBO Post-Baccalaureate Fellowship. The National Radio Astronomy Observatory is a facility of the National Science Foundation operated under cooperative agreement by Associated Universities, Inc. This research was carried out in part at the Jet Propulsion Laboratory, California Institute of Technology, under a contract with the National Aeronautics and Space Administration (80NM0018D0004).  E.F.-J.A. acknowledge support from UNAM-PAPIIT projects IA102023 and IA104725, and from CONAHCyT Ciencia de Frontera project ID: CF-2023-I-506.
\end{acknowledgments}

\clearpage

\bibliography{main.bib}

\appendix
\section{Supplementary Data} \label{app:appendix}
\restartappendixnumbering 

In the following appendix, we list the best-fit amplitudes for the synchrotron and free-free emission models for both MCMC fitting models and the dust emission model for the MCMC fitting model without dust emission and 90\,GHz data (Table \ref{tab:a1}).

\startlongtable
\begin{deluxetable*}{l|ccc|cc}
\label{tab:a1}
\tablecolumns{6}
\tablecaption{Ancillary Derived Parameters at 10\arcsec Angular Resolution}
\tablehead{
\colhead{Source ID} & 
\multicolumn{3}{c}{MCMC w/ dust \& 90\,GHz} &
\multicolumn{2}{c}{MCMC w/o dust \& 90\,GHz}\\
\colhead{} &
\colhead{$A_{s}$} &
\colhead{$A_{ff}$} & 
\colhead{$A_{d}$} & 
\colhead{$A_{s}$} & 
\colhead{$A_{ff}$}
}
\startdata
\cutinhead{Star-Forming Regions}
NGC\,0337\,a & ${0.20}^{0.12}_{-0.07}$ & ${0.56}^{0.12}_{-0.16}$ & ${7.32}^{8.21}_{-5.19}$ & ${0.17}^{0.11}_{-0.07}$ & ${0.64}^{0.13}_{-0.16}$ \\
NGC\,0337\,b & ${0.73}^{0.38}_{-0.24}$ & ${1.00}^{0.28}_{-0.41}$ & ${26.48}^{13.42}_{-12.77}$ & ${0.73}^{0.37}_{-0.24}$ & ${1.00}^{0.29}_{-0.40}$ \\
NGC\,0337\,c & ${0.08}^{0.05}_{-0.04}$ & ${0.21}^{0.10}_{-0.11}$ & ${5.28}^{6.63}_{-3.84}$ & ${0.05}^{0.05}_{-0.03}$ & ${0.31}^{0.11}_{-0.13}$ \\
NGC\,0337\,d & ${0.05}^{0.05}_{-0.03}$ & ${0.44}^{0.09}_{-0.11}$ & ${4.46}^{6.03}_{-3.25}$ & ${0.03}^{0.04}_{-0.02}$ & ${0.55}^{0.09}_{-0.10}$ \\
NGC\,0628\,Enuc.\,4 & ${0.04}^{0.04}_{-0.02}$ & ${0.16}^{0.07}_{-0.07}$ & ${2.98}^{4.26}_{-2.22}$ & ${0.02}^{0.03}_{-0.02}$ & ${0.23}^{0.06}_{-0.07}$ \\
NGC\,0628\,Enuc.\,2 & ${0.05}^{0.04}_{-0.03}$ & ${0.26}^{0.09}_{-0.10}$ & ${7.98}^{8.39}_{-5.55}$ & ${0.04}^{0.04}_{-0.02}$ & ${0.31}^{0.09}_{-0.10}$ \\
NGC\,0628\,Enuc.\,3 & ${0.05}^{0.04}_{-0.03}$ & ${0.41}^{0.09}_{-0.10}$ & ${3.55}^{5.04}_{-2.61}$ & ${0.02}^{0.03}_{-0.02}$ & ${0.52}^{0.08}_{-0.09}$ \\
NGC\,0628\, & ${0.01}^{0.01}_{-0.01}$ & ${0.02}^{0.03}_{-0.02}$ & ${3.19}^{4.09}_{-2.31}$ & ${0.01}^{0.01}_{-0.00}$ & ${0.03}^{0.04}_{-0.02}$ \\
NGC\,0628\,Enuc.\,1 & ${0.15}^{0.08}_{-0.05}$ & ${0.30}^{0.10}_{-0.12}$ & ${2.98}^{4.52}_{-2.21}$ & ${0.08}^{0.06}_{-0.04}$ & ${0.53}^{0.12}_{-0.13}$ \\
NGC\,0925\, & ${0.02}^{0.02}_{-0.01}$ & ${0.16}^{0.06}_{-0.07}$ & ${20.64}^{16.11}_{-13.17}$ & ${0.02}^{0.02}_{-0.01}$ & ${0.17}^{0.06}_{-0.07}$ \\
NGC\,2146\,a & ${6.71}^{1.54}_{-1.67}$ & ${2.31}^{1.67}_{-1.50}$ & ${127.32}^{44.08}_{-53.17}$ & ${6.71}^{1.54}_{-1.71}$ & ${2.35}^{1.69}_{-1.51}$ \\
NGC\,2146\,b & ${58.31}^{2.79}_{-3.14}$ & ${4.10}^{2.81}_{-2.62}$ & ${44.07}^{43.59}_{-30.10}$ & ${34.76}^{14.47}_{-10.19}$ & ${27.25}^{9.83}_{-14.12}$ \\
NGC\,2146\,c & ${19.12}^{0.19}_{-0.24}$ & ${0.08}^{0.14}_{-0.06}$ & ${1.43}^{2.51}_{-1.08}$ & ${12.91}^{5.29}_{-3.79}$ & ${10.15}^{3.72}_{-5.12}$ \\
NGC\,2798\, & ${6.19}^{0.11}_{-0.15}$ & ${0.04}^{0.10}_{-0.03}$ & ${167.34}^{10.26}_{-13.56}$ & ${6.23}^{0.10}_{-0.11}$ & ${0.04}^{0.06}_{-0.03}$ \\
NGC\,2841\, & ${0.13}^{0.07}_{-0.04}$ & ${0.93}^{0.07}_{-0.08}$ & ${106.70}^{10.12}_{-10.16}$ & ${0.12}^{0.07}_{-0.04}$ & ${0.98}^{0.06}_{-0.08}$ \\
NGC\,3049\, & ${0.59}^{0.29}_{-0.20}$ & ${0.62}^{0.21}_{-0.30}$ & ${2.73}^{4.16}_{-2.03}$ & ${0.33}^{0.18}_{-0.11}$ & ${0.97}^{0.14}_{-0.21}$ \\
NGC\,3190\, & ${0.44}^{0.14}_{-0.13}$ & ${0.26}^{0.14}_{-0.15}$ & ${6.84}^{7.88}_{-4.83}$ & ${0.41}^{0.16}_{-0.12}$ & ${0.32}^{0.15}_{-0.18}$ \\
NGC\,3184\, & ${0.10}^{0.05}_{-0.03}$ & ${0.18}^{0.06}_{-0.07}$ & ${54.22}^{12.12}_{-12.13}$ & ${0.10}^{0.05}_{-0.03}$ & ${0.20}^{0.06}_{-0.07}$ \\
NGC\,3198\, & ${0.21}^{0.10}_{-0.07}$ & ${0.23}^{0.10}_{-0.12}$ & ${32.04}^{14.44}_{-14.01}$ & ${0.21}^{0.10}_{-0.07}$ & ${0.23}^{0.10}_{-0.12}$ \\
NGC\,3351\,a & ${2.16}^{0.50}_{-0.45}$ & ${1.02}^{0.49}_{-0.55}$ & ${19.77}^{16.56}_{-13.77}$ & ${1.99}^{0.58}_{-0.50}$ & ${1.21}^{0.58}_{-0.64}$ \\
NGC\,3351\,b & ${3.11}^{0.15}_{-0.24}$ & ${0.18}^{0.24}_{-0.13}$ & ${2.87}^{4.28}_{-2.13}$ & ${2.18}^{0.63}_{-0.53}$ & ${1.34}^{0.61}_{-0.69}$ \\
NGC\,3521\,Enuc.\,1 & ${0.01}^{0.02}_{-0.01}$ & ${0.11}^{0.07}_{-0.06}$ & ${9.35}^{10.13}_{-6.60}$ & ${0.01}^{0.02}_{-0.01}$ & ${0.14}^{0.07}_{-0.07}$ \\
NGC\,3521\,Enuc.\,3 & ${0.02}^{0.02}_{-0.01}$ & ${0.05}^{0.04}_{-0.04}$ & ${23.77}^{15.18}_{-13.50}$ & ${0.02}^{0.02}_{-0.01}$ & ${0.06}^{0.05}_{-0.04}$ \\
NGC\,3521\, & ${0.01}^{0.01}_{-0.00}$ & ${0.02}^{0.03}_{-0.01}$ & ${32.22}^{10.76}_{-10.65}$ & ${0.01}^{0.01}_{-0.00}$ & ${0.02}^{0.03}_{-0.01}$ \\
NGC\,3521\,Enuc.\,2a & ${0.12}^{0.04}_{-0.04}$ & ${0.08}^{0.08}_{-0.06}$ & ${33.01}^{11.47}_{-11.39}$ & ${0.12}^{0.04}_{-0.04}$ & ${0.08}^{0.08}_{-0.06}$ \\
NGC\,3521\,Enuc.\,2b & ${0.03}^{0.03}_{-0.02}$ & ${0.08}^{0.06}_{-0.05}$ & ${6.92}^{7.96}_{-4.94}$ & ${0.03}^{0.03}_{-0.02}$ & ${0.10}^{0.07}_{-0.06}$ \\
NGC\,3627\, & ${1.36}^{0.50}_{-0.39}$ & ${0.96}^{0.44}_{-0.52}$ & ${44.98}^{45.89}_{-31.34}$ & ${1.30}^{0.48}_{-0.37}$ & ${1.06}^{0.41}_{-0.51}$ \\
NGC\,3627\,Enuc.\,1 & ${0.07}^{0.06}_{-0.04}$ & ${1.99}^{0.12}_{-0.13}$ & ${37.42}^{43.98}_{-26.65}$ & ${0.07}^{0.06}_{-0.04}$ & ${2.02}^{0.12}_{-0.13}$ \\
NGC\,3627\,Enuc.\,2 & ${1.31}^{0.67}_{-0.41}$ & ${4.11}^{0.43}_{-0.68}$ & ${156.08}^{60.39}_{-60.77}$ & ${1.27}^{0.68}_{-0.40}$ & ${4.21}^{0.42}_{-0.68}$ \\
NGC\,3773\, & ${0.21}^{0.10}_{-0.06}$ & ${0.87}^{0.10}_{-0.12}$ & ${21.66}^{10.63}_{-10.25}$ & ${0.20}^{0.10}_{-0.06}$ & ${0.88}^{0.10}_{-0.13}$ \\
NGC\,3938\,b & ${0.00}^{0.01}_{-0.00}$ & ${0.03}^{0.03}_{-0.02}$ & ${16.63}^{11.36}_{-9.79}$ & ${0.00}^{0.01}_{-0.00}$ & ${0.03}^{0.03}_{-0.02}$ \\
NGC\,3938\,a & ${0.00}^{0.01}_{-0.00}$ & ${0.01}^{0.01}_{-0.01}$ & ${10.77}^{10.13}_{-7.28}$ & ${0.00}^{0.01}_{-0.00}$ & ${0.01}^{0.02}_{-0.01}$ \\
NGC\,3938\,Enuc.\,2a & ${0.02}^{0.02}_{-0.01}$ & ${0.11}^{0.05}_{-0.05}$ & ${13.01}^{13.62}_{-8.92}$ & ${0.01}^{0.02}_{-0.01}$ & ${0.12}^{0.05}_{-0.05}$ \\
NGC\,3938\,Enuc.\,2b & ${0.02}^{0.02}_{-0.01}$ & ${0.19}^{0.05}_{-0.06}$ & ${4.95}^{7.05}_{-3.67}$ & ${0.01}^{0.02}_{-0.01}$ & ${0.22}^{0.05}_{-0.05}$ \\
NGC\,4254\,Enuc.\,2a & ${0.01}^{0.02}_{-0.01}$ & ${0.25}^{0.05}_{-0.05}$ & ${2.70}^{3.78}_{-1.98}$ & ${0.01}^{0.01}_{-0.01}$ & ${0.30}^{0.06}_{-0.06}$ \\
NGC\,4254\,Enuc.\,2b & ${0.01}^{0.02}_{-0.01}$ & ${0.19}^{0.05}_{-0.05}$ & ${3.04}^{4.07}_{-2.23}$ & ${0.01}^{0.01}_{-0.01}$ & ${0.24}^{0.05}_{-0.06}$ \\
NGC\,4254\,a & ${0.09}^{0.03}_{-0.03}$ & ${0.04}^{0.04}_{-0.03}$ & ${26.72}^{6.94}_{-7.01}$ & ${0.09}^{0.03}_{-0.03}$ & ${0.04}^{0.05}_{-0.03}$ \\
NGC\,4254\,b & ${0.01}^{0.02}_{-0.01}$ & ${0.41}^{0.06}_{-0.07}$ & ${57.70}^{7.99}_{-8.09}$ & ${0.01}^{0.02}_{-0.01}$ & ${0.42}^{0.06}_{-0.07}$ \\
NGC\,4254\,Enuc.\,1b & ${0.06}^{0.05}_{-0.03}$ & ${0.23}^{0.08}_{-0.09}$ & ${3.12}^{4.45}_{-2.29}$ & ${0.03}^{0.04}_{-0.02}$ & ${0.36}^{0.09}_{-0.10}$ \\
NGC\,4254\,c & ${0.05}^{0.03}_{-0.03}$ & ${0.07}^{0.06}_{-0.05}$ & ${71.13}^{7.04}_{-7.28}$ & ${0.05}^{0.03}_{-0.03}$ & ${0.08}^{0.06}_{-0.05}$ \\
NGC\,4254\,Enuc.\,1c & ${0.13}^{0.05}_{-0.04}$ & ${0.14}^{0.09}_{-0.08}$ & ${8.61}^{7.40}_{-5.63}$ & ${0.12}^{0.06}_{-0.04}$ & ${0.17}^{0.11}_{-0.10}$ \\
NGC\,4254\,d & ${0.12}^{0.05}_{-0.04}$ & ${0.13}^{0.11}_{-0.08}$ & ${43.26}^{8.47}_{-9.08}$ & ${0.12}^{0.05}_{-0.04}$ & ${0.14}^{0.11}_{-0.09}$ \\
NGC\,4254\,e & ${0.02}^{0.03}_{-0.01}$ & ${0.15}^{0.08}_{-0.08}$ & ${18.11}^{8.05}_{-7.99}$ & ${0.02}^{0.03}_{-0.01}$ & ${0.15}^{0.08}_{-0.08}$ \\
NGC\,4254\,f & ${0.02}^{0.02}_{-0.01}$ & ${0.05}^{0.05}_{-0.03}$ & ${12.10}^{6.78}_{-6.41}$ & ${0.02}^{0.02}_{-0.01}$ & ${0.05}^{0.06}_{-0.04}$ \\
NGC\,4321\,Enuc.\,2a & ${0.02}^{0.02}_{-0.01}$ & ${0.08}^{0.05}_{-0.05}$ & ${7.09}^{9.17}_{-5.17}$ & ${0.02}^{0.02}_{-0.01}$ & ${0.10}^{0.05}_{-0.05}$ \\
NGC\,4321\,Enuc.\,2b & ${0.03}^{0.03}_{-0.02}$ & ${0.07}^{0.05}_{-0.05}$ & ${9.55}^{10.47}_{-6.70}$ & ${0.03}^{0.03}_{-0.02}$ & ${0.08}^{0.06}_{-0.05}$ \\
NGC\,4321\,Enuc.\,2 & ${0.02}^{0.02}_{-0.01}$ & ${0.05}^{0.04}_{-0.03}$ & ${9.78}^{10.48}_{-6.86}$ & ${0.02}^{0.02}_{-0.01}$ & ${0.06}^{0.05}_{-0.04}$ \\
NGC\,4321\,a & ${2.77}^{0.45}_{-0.60}$ & ${0.71}^{0.64}_{-0.48}$ & ${49.05}^{18.69}_{-21.11}$ & ${2.75}^{0.48}_{-0.59}$ & ${0.73}^{0.63}_{-0.50}$ \\
NGC\,4321\,b & ${3.12}^{0.48}_{-0.68}$ & ${0.75}^{0.74}_{-0.51}$ & ${77.91}^{19.60}_{-24.00}$ & ${3.17}^{0.47}_{-0.62}$ & ${0.71}^{0.66}_{-0.49}$ \\
NGC\,4321\,Enuc.\,1 & ${0.02}^{0.02}_{-0.01}$ & ${0.03}^{0.04}_{-0.02}$ & ${16.45}^{12.00}_{-9.99}$ & ${0.02}^{0.02}_{-0.01}$ & ${0.03}^{0.04}_{-0.02}$ \\
NGC\,4536\, & ${12.65}^{0.78}_{-0.85}$ & ${2.44}^{0.96}_{-0.87}$ & ${253.32}^{26.48}_{-31.06}$ & ${12.31}^{0.68}_{-0.63}$ & ${2.86}^{0.71}_{-0.75}$ \\
NGC\,4559\,a & ${0.01}^{0.01}_{-0.01}$ & ${0.13}^{0.04}_{-0.04}$ & ${23.33}^{12.43}_{-11.72}$ & ${0.01}^{0.01}_{-0.01}$ & ${0.13}^{0.04}_{-0.04}$ \\
NGC\,4559\,b & ${0.01}^{0.02}_{-0.01}$ & ${0.11}^{0.04}_{-0.04}$ & ${28.89}^{12.70}_{-12.42}$ & ${0.01}^{0.01}_{-0.01}$ & ${0.12}^{0.04}_{-0.04}$ \\
NGC\,4559\,c & ${0.01}^{0.02}_{-0.01}$ & ${0.19}^{0.04}_{-0.04}$ & ${29.46}^{12.60}_{-12.52}$ & ${0.01}^{0.02}_{-0.01}$ & ${0.20}^{0.04}_{-0.04}$ \\
NGC\,4569\, & ${1.10}^{0.27}_{-0.29}$ & ${0.43}^{0.32}_{-0.28}$ & ${63.96}^{16.91}_{-17.32}$ & ${1.11}^{0.28}_{-0.28}$ & ${0.43}^{0.31}_{-0.28}$ \\
NGC\,4579\, & ${46.45}^{5.67}_{-6.84}$ & ${8.64}^{7.04}_{-5.77}$ & ${3245.77}^{104.91}_{-207.57}$ & ${40.68}^{4.55}_{-3.27}$ & ${15.52}^{3.34}_{-4.66}$ \\
NGC\,4594\,a & ${77.21}^{0.66}_{-1.17}$ & ${0.90}^{1.17}_{-0.65}$ & ${2060.64}^{14.21}_{-14.91}$ & ${77.59}^{0.54}_{-0.98}$ & ${0.73}^{0.98}_{-0.53}$ \\
NGC\,4625\, & ${0.00}^{0.01}_{-0.00}$ & ${0.09}^{0.03}_{-0.03}$ & ${21.08}^{26.86}_{-15.23}$ & ${0.00}^{0.01}_{-0.00}$ & ${0.10}^{0.03}_{-0.03}$ \\
NGC\,4631\,a & ${1.35}^{0.62}_{-0.44}$ & ${1.55}^{0.52}_{-0.67}$ & ${63.85}^{19.74}_{-19.84}$ & ${1.33}^{0.59}_{-0.40}$ & ${1.58}^{0.46}_{-0.64}$ \\
NGC\,4631\,b & ${1.68}^{0.87}_{-0.54}$ & ${2.29}^{0.58}_{-0.93}$ & ${32.34}^{19.45}_{-17.99}$ & ${1.68}^{0.89}_{-0.56}$ & ${2.29}^{0.62}_{-0.93}$ \\
NGC\,4631\,c & ${0.54}^{0.28}_{-0.17}$ & ${3.84}^{0.24}_{-0.33}$ & ${66.62}^{15.60}_{-15.12}$ & ${0.53}^{0.26}_{-0.16}$ & ${3.87}^{0.23}_{-0.31}$ \\
NGC\,4631\,d & ${0.06}^{0.05}_{-0.03}$ & ${0.97}^{0.13}_{-0.15}$ & ${60.37}^{13.25}_{-12.82}$ & ${0.05}^{0.05}_{-0.03}$ & ${0.99}^{0.12}_{-0.15}$ \\
NGC\,4631\,e & ${0.17}^{0.10}_{-0.06}$ & ${1.77}^{0.16}_{-0.17}$ & ${88.53}^{13.68}_{-13.43}$ & ${0.16}^{0.09}_{-0.06}$ & ${1.81}^{0.15}_{-0.16}$ \\
NGC\,4631\,Enuc.\,2b & ${0.03}^{0.03}_{-0.02}$ & ${1.03}^{0.07}_{-0.08}$ & ${6.53}^{8.44}_{-4.74}$ & ${0.02}^{0.03}_{-0.01}$ & ${1.07}^{0.06}_{-0.07}$ \\
NGC\,4725\,a & ${0.00}^{0.01}_{-0.00}$ & ${0.24}^{0.03}_{-0.03}$ & ${8.56}^{7.84}_{-5.76}$ & ${0.00}^{0.01}_{-0.00}$ & ${0.25}^{0.03}_{-0.03}$ \\
NGC\,5194\,Enuc.\,6a & ${0.02}^{0.03}_{-0.02}$ & ${0.57}^{0.04}_{-0.05}$ & ${11.24}^{14.98}_{-8.17}$ & ${0.02}^{0.03}_{-0.02}$ & ${0.58}^{0.04}_{-0.05}$ \\
NGC\,5194\,Enuc.\,3 & ${0.02}^{0.04}_{-0.02}$ & ${0.66}^{0.06}_{-0.07}$ & ${8.27}^{10.83}_{-6.05}$ & ${0.02}^{0.03}_{-0.02}$ & ${0.68}^{0.06}_{-0.06}$ \\
NGC\,5194\,Enuc.\,11a & ${0.09}^{0.05}_{-0.04}$ & ${0.11}^{0.09}_{-0.07}$ & ${8.83}^{11.17}_{-6.35}$ & ${0.09}^{0.05}_{-0.04}$ & ${0.14}^{0.09}_{-0.09}$ \\
NGC\,5194\,Enuc.\,11b & ${0.03}^{0.04}_{-0.02}$ & ${0.14}^{0.07}_{-0.07}$ & ${29.84}^{17.76}_{-16.30}$ & ${0.03}^{0.04}_{-0.02}$ & ${0.15}^{0.07}_{-0.07}$ \\
NGC\,5194\,Enuc.\,11d & ${0.02}^{0.02}_{-0.01}$ & ${0.06}^{0.05}_{-0.04}$ & ${7.76}^{10.16}_{-5.73}$ & ${0.02}^{0.03}_{-0.01}$ & ${0.08}^{0.05}_{-0.05}$ \\
NGC\,5194\,Enuc.\,11c & ${0.16}^{0.05}_{-0.04}$ & ${0.05}^{0.06}_{-0.04}$ & ${9.74}^{12.17}_{-7.03}$ & ${0.16}^{0.05}_{-0.04}$ & ${0.06}^{0.07}_{-0.05}$ \\
NGC\,5194\,c & ${0.46}^{0.13}_{-0.12}$ & ${0.19}^{0.17}_{-0.13}$ & ${26.68}^{16.52}_{-14.97}$ & ${0.45}^{0.13}_{-0.12}$ & ${0.20}^{0.17}_{-0.14}$ \\
NGC\,5194\,Enuc.\,11e & ${0.06}^{0.06}_{-0.04}$ & ${0.28}^{0.08}_{-0.09}$ & ${6.60}^{9.24}_{-4.84}$ & ${0.05}^{0.05}_{-0.03}$ & ${0.33}^{0.08}_{-0.09}$ \\
NGC\,5194\,b & ${0.61}^{0.13}_{-0.14}$ & ${0.16}^{0.16}_{-0.11}$ & ${8.59}^{10.53}_{-6.19}$ & ${0.59}^{0.15}_{-0.15}$ & ${0.23}^{0.19}_{-0.16}$ \\
NGC\,5194\,e & ${2.14}^{0.12}_{-0.15}$ & ${0.10}^{0.15}_{-0.07}$ & ${19.20}^{14.97}_{-12.06}$ & ${2.15}^{0.12}_{-0.15}$ & ${0.10}^{0.15}_{-0.07}$ \\
NGC\,5194\,d & ${1.95}^{0.10}_{-0.13}$ & ${0.08}^{0.12}_{-0.06}$ & ${61.11}^{16.45}_{-16.86}$ & ${1.97}^{0.10}_{-0.12}$ & ${0.07}^{0.11}_{-0.05}$ \\
NGC\,5194\,Enuc.\,1c & ${0.07}^{0.06}_{-0.04}$ & ${0.27}^{0.10}_{-0.11}$ & ${13.90}^{13.95}_{-9.63}$ & ${0.06}^{0.06}_{-0.04}$ & ${0.31}^{0.10}_{-0.11}$ \\
NGC\,5194\,Enuc.\,4a & ${0.01}^{0.02}_{-0.01}$ & ${0.12}^{0.05}_{-0.05}$ & ${31.24}^{20.14}_{-17.65}$ & ${0.01}^{0.02}_{-0.01}$ & ${0.13}^{0.05}_{-0.05}$ \\
NGC\,5194\,Enuc.\,10a & ${0.08}^{0.05}_{-0.04}$ & ${0.12}^{0.09}_{-0.08}$ & ${5.21}^{7.29}_{-3.81}$ & ${0.06}^{0.05}_{-0.04}$ & ${0.20}^{0.10}_{-0.10}$ \\
NGC\,5194\,Enuc.\,4c & ${0.08}^{0.05}_{-0.04}$ & ${0.19}^{0.07}_{-0.07}$ & ${26.48}^{18.61}_{-15.98}$ & ${0.07}^{0.05}_{-0.04}$ & ${0.20}^{0.06}_{-0.07}$ \\
NGC\,5194\,a & ${0.09}^{0.08}_{-0.05}$ & ${0.84}^{0.14}_{-0.16}$ & ${10.74}^{12.48}_{-7.60}$ & ${0.07}^{0.07}_{-0.05}$ & ${0.93}^{0.14}_{-0.16}$ \\
NGC\,5194\,Enuc.\,10b & ${0.10}^{0.08}_{-0.05}$ & ${0.29}^{0.11}_{-0.12}$ & ${5.34}^{7.64}_{-3.95}$ & ${0.07}^{0.07}_{-0.04}$ & ${0.39}^{0.10}_{-0.12}$ \\
NGC\,5194\,Enuc.\,4d & ${0.08}^{0.05}_{-0.04}$ & ${0.13}^{0.07}_{-0.07}$ & ${33.72}^{22.75}_{-19.88}$ & ${0.07}^{0.05}_{-0.04}$ & ${0.14}^{0.07}_{-0.07}$ \\
NGC\,5194\,Enuc.\,5a & ${0.03}^{0.04}_{-0.02}$ & ${0.40}^{0.07}_{-0.08}$ & ${32.34}^{22.42}_{-19.03}$ & ${0.02}^{0.03}_{-0.02}$ & ${0.42}^{0.07}_{-0.08}$ \\
NGC\,5194\,Enuc.\,9 & ${0.03}^{0.04}_{-0.02}$ & ${0.24}^{0.07}_{-0.08}$ & ${4.23}^{6.27}_{-3.13}$ & ${0.02}^{0.03}_{-0.02}$ & ${0.30}^{0.07}_{-0.07}$ \\
NGC\,5194\,Enuc.\,7a & ${0.05}^{0.05}_{-0.03}$ & ${0.15}^{0.07}_{-0.07}$ & ${5.73}^{8.38}_{-4.24}$ & ${0.04}^{0.04}_{-0.02}$ & ${0.18}^{0.07}_{-0.07}$ \\
NGC\,5194\,Enuc.\,8 & ${0.13}^{0.10}_{-0.06}$ & ${0.52}^{0.09}_{-0.12}$ & ${9.09}^{11.30}_{-6.55}$ & ${0.12}^{0.09}_{-0.05}$ & ${0.55}^{0.09}_{-0.11}$ \\
NGC\,5194\,Enuc.\,7b & ${0.20}^{0.10}_{-0.07}$ & ${0.28}^{0.09}_{-0.12}$ & ${6.57}^{9.25}_{-4.84}$ & ${0.18}^{0.09}_{-0.06}$ & ${0.33}^{0.09}_{-0.11}$ \\
NGC\,5194\,Enuc.\,7c & ${0.07}^{0.04}_{-0.03}$ & ${0.07}^{0.06}_{-0.05}$ & ${17.28}^{17.39}_{-11.86}$ & ${0.07}^{0.04}_{-0.03}$ & ${0.08}^{0.06}_{-0.05}$ \\
NGC\,5474\, & ${0.01}^{0.01}_{-0.01}$ & ${0.03}^{0.03}_{-0.02}$ & ${9.18}^{12.12}_{-6.73}$ & ${0.01}^{0.01}_{-0.01}$ & ${0.03}^{0.03}_{-0.02}$ \\
NGC\,5713\,Enuc.\,2a & ${0.72}^{0.32}_{-0.22}$ & ${0.77}^{0.27}_{-0.34}$ & ${89.36}^{18.49}_{-17.65}$ & ${0.70}^{0.32}_{-0.23}$ & ${0.81}^{0.27}_{-0.35}$ \\
NGC\,5713\,Enuc.\,2b & ${0.23}^{0.07}_{-0.06}$ & ${0.08}^{0.10}_{-0.06}$ & ${76.93}^{15.05}_{-15.58}$ & ${0.24}^{0.07}_{-0.06}$ & ${0.09}^{0.10}_{-0.07}$ \\
NGC\,5713\, & ${3.26}^{0.74}_{-1.13}$ & ${1.14}^{1.15}_{-0.75}$ & ${145.67}^{21.80}_{-30.13}$ & ${3.42}^{0.69}_{-0.83}$ & ${1.03}^{0.84}_{-0.69}$ \\
NGC\,5713\,Enuc.\,1 & ${4.51}^{0.44}_{-0.97}$ & ${0.60}^{0.99}_{-0.45}$ & ${93.85}^{19.30}_{-27.13}$ & ${4.64}^{0.37}_{-0.63}$ & ${0.50}^{0.64}_{-0.37}$ \\
NGC\,6946\,Enuc.\,4c & ${0.15}^{0.12}_{-0.07}$ & ${1.48}^{0.11}_{-0.15}$ & ${64.75}^{17.18}_{-16.99}$ & ${0.15}^{0.11}_{-0.07}$ & ${1.50}^{0.11}_{-0.14}$ \\
NGC\,6946\,Enuc.\,8 & ${0.22}^{0.09}_{-0.06}$ & ${1.02}^{0.09}_{-0.11}$ & ${31.59}^{15.10}_{-14.62}$ & ${0.21}^{0.08}_{-0.06}$ & ${1.04}^{0.09}_{-0.11}$ \\
NGC\,6946\,Enuc.\,5a & ${0.01}^{0.01}_{-0.01}$ & ${0.03}^{0.03}_{-0.02}$ & ${14.71}^{13.21}_{-9.76}$ & ${0.01}^{0.01}_{-0.01}$ & ${0.03}^{0.03}_{-0.02}$ \\
NGC\,6946\,Enuc.\,5b & ${0.02}^{0.03}_{-0.02}$ & ${0.37}^{0.04}_{-0.05}$ & ${25.12}^{15.19}_{-13.82}$ & ${0.02}^{0.03}_{-0.01}$ & ${0.38}^{0.04}_{-0.05}$ \\
NGC\,6946\,Enuc.\,3a & ${0.01}^{0.02}_{-0.01}$ & ${0.28}^{0.04}_{-0.04}$ & ${28.11}^{15.64}_{-14.46}$ & ${0.01}^{0.02}_{-0.01}$ & ${0.29}^{0.04}_{-0.04}$ \\
NGC\,6946\,Enuc.\,3b & ${0.01}^{0.02}_{-0.01}$ & ${0.58}^{0.04}_{-0.04}$ & ${40.43}^{15.86}_{-15.60}$ & ${0.01}^{0.02}_{-0.01}$ & ${0.59}^{0.04}_{-0.04}$ \\
NGC\,6946\,b & ${14.00}^{0.33}_{-1.03}$ & ${0.47}^{1.10}_{-0.35}$ & ${345.56}^{16.06}_{-24.90}$ & ${14.09}^{0.29}_{-0.57}$ & ${0.40}^{0.61}_{-0.29}$ \\
NGC\,6946\,Enuc.\,6b & ${0.05}^{0.05}_{-0.03}$ & ${0.47}^{0.08}_{-0.10}$ & ${22.88}^{15.25}_{-13.41}$ & ${0.05}^{0.05}_{-0.03}$ & ${0.48}^{0.08}_{-0.09}$ \\
NGC\,6946\,Enuc.\,9 & ${0.01}^{0.02}_{-0.01}$ & ${1.69}^{0.05}_{-0.06}$ & ${65.24}^{14.17}_{-14.44}$ & ${0.01}^{0.02}_{-0.01}$ & ${1.70}^{0.05}_{-0.06}$ \\
NGC\,6946\,Enuc.\,7 & ${0.01}^{0.01}_{-0.00}$ & ${1.09}^{0.05}_{-0.05}$ & ${19.94}^{13.48}_{-11.77}$ & ${0.01}^{0.01}_{-0.00}$ & ${1.10}^{0.05}_{-0.05}$ \\
NGC\,6946\,Enuc.\,1 & ${0.03}^{0.04}_{-0.02}$ & ${0.55}^{0.05}_{-0.06}$ & ${7.69}^{10.05}_{-5.52}$ & ${0.03}^{0.03}_{-0.02}$ & ${0.56}^{0.05}_{-0.06}$ \\
NGC\,7331\, & ${0.01}^{0.01}_{-0.01}$ & ${0.03}^{0.04}_{-0.02}$ & ${40.57}^{9.70}_{-9.65}$ & ${0.01}^{0.01}_{-0.01}$ & ${0.03}^{0.04}_{-0.02}$ \\
\cutinhead{Likely Associated with Supernovae}
NGC\,6946\,Enuc.\,6a & ${0.09}^{0.06}_{-0.04}$ & ${1.07}^{0.09}_{-0.10}$ & ${62.20}^{16.41}_{-16.32}$ & ${0.08}^{0.06}_{-0.04}$ & ${1.09}^{0.09}_{-0.10}$ \\
\cutinhead{Likely AME Candidates}
NGC\,4254\,Enuc.\,1a & ${0.02}^{0.03}_{-0.01}$ & ${0.16}^{0.06}_{-0.07}$ & ${4.76}^{5.53}_{-3.39}$ & ${0.02}^{0.02}_{-0.01}$ & ${0.20}^{0.07}_{-0.07}$ \\
NGC\,4725\,b & ${0.01}^{0.01}_{-0.00}$ & ${0.23}^{0.03}_{-0.03}$ & ${8.64}^{8.17}_{-5.81}$ & ${0.01}^{0.01}_{-0.00}$ & ${0.24}^{0.03}_{-0.04}$ \\
NGC\,5194\,Enuc.\,2 & ${0.09}^{0.07}_{-0.04}$ & ${0.71}^{0.09}_{-0.10}$ & ${9.09}^{11.38}_{-6.52}$ & ${0.08}^{0.06}_{-0.04}$ & ${0.75}^{0.08}_{-0.10}$ \\
NGC\,5194\,Enuc.\,1a & ${0.04}^{0.05}_{-0.03}$ & ${0.73}^{0.11}_{-0.12}$ & ${5.61}^{8.04}_{-4.18}$ & ${0.02}^{0.04}_{-0.02}$ & ${0.85}^{0.11}_{-0.12}$ \\
NGC\,5194\,Enuc.\,1b & ${0.03}^{0.04}_{-0.02}$ & ${0.55}^{0.08}_{-0.09}$ & ${10.15}^{11.67}_{-7.23}$ & ${0.02}^{0.03}_{-0.02}$ & ${0.59}^{0.08}_{-0.09}$ \\
NGC\,5194\,Enuc.\,4b & ${0.05}^{0.05}_{-0.03}$ & ${0.40}^{0.06}_{-0.07}$ & ${13.65}^{15.34}_{-9.57}$ & ${0.05}^{0.05}_{-0.03}$ & ${0.42}^{0.06}_{-0.07}$ \\
NGC\,6946\,Enuc.\,4a & ${0.01}^{0.02}_{-0.01}$ & ${1.86}^{0.04}_{-0.04}$ & ${8.46}^{10.23}_{-5.99}$ & ${0.01}^{0.01}_{-0.00}$ & ${1.88}^{0.04}_{-0.04}$ \\
NGC\,6946\,Enuc.\,4b & ${0.01}^{0.01}_{-0.01}$ & ${0.81}^{0.04}_{-0.04}$ & ${50.56}^{16.21}_{-16.49}$ & ${0.01}^{0.01}_{-0.01}$ & ${0.82}^{0.04}_{-0.04}$ \\
NGC\,6946\,a & ${0.01}^{0.02}_{-0.01}$ & ${1.03}^{0.10}_{-0.10}$ & ${51.09}^{12.95}_{-12.96}$ & ${0.01}^{0.02}_{-0.01}$ & ${1.04}^{0.09}_{-0.10}$ \\
NGC\,6946\,c & ${0.29}^{0.19}_{-0.11}$ & ${1.88}^{0.17}_{-0.24}$ & ${9.09}^{10.40}_{-6.49}$ & ${0.22}^{0.16}_{-0.09}$ & ${2.03}^{0.19}_{-0.24}$ \\
NGC\,6946\,Enuc.\,2a & ${0.01}^{0.02}_{-0.01}$ & ${0.28}^{0.07}_{-0.07}$ & ${83.15}^{16.89}_{-16.51}$ & ${0.01}^{0.02}_{-0.01}$ & ${0.30}^{0.07}_{-0.07}$ \\
NGC\,6946\,Enuc.\,2b & ${0.02}^{0.03}_{-0.01}$ & ${1.74}^{0.07}_{-0.07}$ & ${42.23}^{17.08}_{-16.53}$ & ${0.02}^{0.03}_{-0.01}$ & ${1.75}^{0.07}_{-0.07}$ \\
\enddata
\end{deluxetable*}

\end{document}